\documentclass[lettersize,journal]{IEEEtran}
\usepackage{amsmath,amsfonts}
\usepackage{algorithmic}
\usepackage{algorithm}
\usepackage{array}
\usepackage[caption=false,font=normalsize,labelfont=sf,textfont=sf]{subfig}
\usepackage{textcomp}
\usepackage{stfloats}
\usepackage{url}
\usepackage{verbatim}
\usepackage{graphicx}
\usepackage{cite}
\usepackage{svg}

\usepackage{amsmath,amssymb,amsfonts}
\usepackage{algorithmic}
\usepackage{color}
\usepackage{xcolor}
\usepackage{soul}
\usepackage{multirow}
\usepackage{float}
\usepackage{svg}
\usepackage{xcolor} 
\usepackage{caption}
\usepackage{arydshln}
\usepackage{hyperref}

\newcommand{\thickhline}{%
 \noalign {\ifnum 0=`}\fi \hrule height 1pt
 \futurelet \reserved@a \@xhline
}

\begin{document}

\title{WiFaKey: Generating Cryptographic Keys from Face in the Wild}

\author{Xingbo Dong,~\IEEEmembership{Member, IEEE}, Hui Zhang, Yen Lung Lai, Zhe Jin,~\IEEEmembership{Member, IEEE}, Junduan Huang,\\ Wenxiong Kang,~\IEEEmembership{Member, IEEE} and Andrew Beng Jin Teoh,~\IEEEmembership{Senior Member, IEEE}
\thanks{This work was supported by the Brain Pool Program through the National Research Foundation of Korea (NRF) funded by the Ministry of Science and ICT under Grant 2021H1D3A2A01099396, the NRF grant funded by the Korea government (MSIP) (NO. NRF-2022R1A2C1010710), the Yonsei University Research Fund (Yonsei Frontier Lab. Yonsei Frontier Program for Outstanding Scholars) of 2022.}
\thanks{X. Dong, H. Zhang, Y. L. Lai, and Z. Jin are with Anhui Provincial International Joint Research Center for Advanced Technology in Medical Imaging, Anhui Provincial Key Laboratory of Secure Artificial Intelligence, Anhui University, Hefei 230093, China. This work was partially conducted while Dong was affiliated with Yonsei University.}
\thanks{J. Huang is with the School of Artificial Intelligence, South China Normal University, Foshan, 528225, China.}
\thanks{W. Kang is with the School of Automation Science and Engineering and the School of Future Technology, South China University of Technology, Guangzhou 510641, China, and also with the Pazhou Laboratory, Guangzhou 510335, China.}
\thanks{A. B. J. Teoh is with the School of Electrical and Electronic Engineering, College of Engineering, Yonsei University, Seoul 120749, Republic of Korea.}
\thanks{Corresponding author: A. B. J. Teoh (bjteoh@yonsei.ac.kr).}
}

\markboth{Journal of \LaTeX\ Class Files,~Vol.~14, No.~8, August~2021}%
{Shell \MakeLowercase{\textit{et al.}}: A Sample Article Using IEEEtran.cls for IEEE Journals}


\maketitle

\begin{abstract}
Deriving a unique cryptographic key from biometric measurements is a challenging task due to the existing noise gap between the biometric measurements and error correction coding. Additionally, privacy and security concerns arise as biometric measurements are inherently linked to the user. Bio-cryptosystems represent a key branch of solutions aimed at addressing these issues. However, many existing bio-cryptosystems rely on handcrafted feature extractors and error correction codes (ECC), often leading to performance degradation.
To address these challenges and improve the reliability of biometric measurements, we propose a novel biometric cryptosystem named \textit{WiFaKey}, for generating cryptographic keys from face in unconstrained settings. Specifically,
\textit{WiFaKey} first introduces an adaptive random masking-driven feature transformation pipeline, AdaMTrans. AdaMTrans effectively quantizes and binarizes real-valued features and incorporates an adaptive random masking scheme to align the bit error rate with error correction requirements, thereby mitigating the noise gap. 
Besides, \textit{WiFaKey} incorporates a supervised learning-based neural decoding scheme called Neural-MS decoder, which delivers a more robust error correction performance with less iteration than non-learning decoders, thereby alleviating the performance degradation.
We evaluated \textit{WiFaKey} using widely adopted face feature extractors on six large unconstrained and two constrained datasets. On the LFW dataset, \textit{WiFaKey} achieved an average Genuine Match Rate of 85.45\% and 85.20\% at a 0\% False Match Rate for MagFace and AdaFace features, respectively. Our comprehensive comparative analysis shows a significant performance improvement of \textit{WiFaKey}. The source code of our work is available at \href{https://github.com/xingbod/WiFaKey}{github.com/xingbod/WiFaKey}. 
\end{abstract}


\section{Introduction\label{sec.intro}}

Biometrics offer reliable identity authentication by measuring unique traits like fingerprints, facial features, iris patterns, voice, and gait, converting these into digital data for analysis and comparison \cite{chen2021weakly, ZhangMulti}. However, directly using biometric measurements as a \textbf{cryptographic key} presents several challenges: 1) the variability and error rates in biometric measurements, such as iris codes with an error rate of 10\%–30\% \cite{daugman2009iris} and face features with an error rate of around 30\% (see Section \ref{section.masking}); and 2) the stringent privacy and security policies required by regulations like the General Data Protection Regulation (GDPR) \cite{gdpr}.

Among existing works \cite{rathgeb2011survey,lei2021privface,walia2020design}, biometric cryptosystem (BC) \cite{xi2010bio} is a prominent instrument system devised to generate cryptography keys from biometric measurements. BC can be further classified into key generation and key binding, where key generation BC directly generates a key from the biometric measurements such as fuzzy extractors \cite{dodis2004fuzzy}, and key binding BC binds a key to a biometric datum such as fuzzy commitments\cite{Juels-CCCS-fuzzycommitment-1999} and fuzzy vaults\cite{lai2018secure}. Helper or auxiliary data is also generated by the BC and stored in the database simultaneously.
Cryptographically, there should be minimal information leakage about original biometric measurements and keys from the helper data, as retrieving the key or the original biometric measurements would be computationally difficult based on the helper data. Instrumentally, biometric measurements of the correct person are required before the key can be released or reconstructed in BC.

Fuzzy commitment \cite{Juels-CCCS-fuzzycommitment-1999} is a typical representative BC scheme used in biometric systems to store and retrieve biometric measurements securely, which has been deployed with a variety of biometric modalities, including iris~\cite{hao-TC-irisfuzzycommit-2006}, face~\cite{kelkboom-ICB-facefuzzycommit-2007}, fingerprints \cite{li2012effective}, etc. It generates a codeword from an Error Correction Code (ECC) and computes the difference between the biometric vector and the codeword, which is hashed and stored as a commitment. In the de-commitment step, the user provides their biometric measurements, and a codeword is generated by XOR operation on biometric measurements with the commitment difference. If the distance between the stored and input biometric measurements is below a certain threshold, the original codeword can be restored using the ECC decoder. However, when encountering unconstrained scenarios \cite{sellahewa2010image,rai2022robust} in the wild, there remain several obstacles to fuzzy commitment:
\begin{itemize}
\item \textbf{Performance Degradation.} The fuzzy commitment scheme, which is typically applied to binary biometric measurements such as IrisCode \cite{daugman2006probing}, necessitates the conversion of real-valued biometric features into binary vectors. This conversion process inherently involves a loss of information, resulting in degraded performance \cite{rathgeb2022deep, osadchy2018all}. Furthermore, most fuzzy commitment schemes are designed and evaluated using controlled biometric datasets, which assume cooperative authentication scenarios \cite{rathgeb2011survey}. However, in unconstrained environments, where biometric measurements are subject to variations and noise, significant performance degradation is observed \cite{bringer2007optimal}.  \textbf{Thus, when the fuzzy commitment scheme is deployed in unconstrained scenarios, the primary issue becomes measurement-induced performance degradation.}

\item \textbf{Noise Gap.} The gap between the biometric measurement noise level and the ECC's error tolerance capability is a critical issue when using fuzzy commitment in biometrics. The ideal scenario would feature zero intra-class variation and maximum inter-class variation in the extracted biometric features, ensuring precise and accurate measurements. In this perfect scenario, the ECC would provide 100\% error correction capability for the same user despite any noise in the measurements, while achieving 0\% error correction for different users. However, this level of performance is practically unattainable due to inherent noise and variability in biometric measurements. For instance, in the case of binarized face features, intra-class samples often exhibit an error rate of approximately 30\% (see Section \ref{section.masking}). Given a BCH code with parameters ($n,k,t$), where $n=511$, the maximum number of correctable error bits $t$ is 121 bits, and the corresponding secret key length $k$ is only 10 bits, resulting in low secrecy utility. Alternatively, the error rate can be artificially increased by interleaving the binary vector with a constant binary vector, \textbf{but this significantly increases the processing cost due to the multiple encoding-decoding steps required}, which is impractical for real-time biometric authentication systems.
\end{itemize}

In this study, we propose a neural error-decoder-based bio-cryptosystem based on face, named \textit{WiFaKey}, specifically designed for handling biometric measurements under unconstrained conditions. \textit{WiFaKey} first employs an adaptive random masking-driven feature transformation pipeline, \textit{AdaMTrans}. This pipeline quantizes and binarizes real-valued features extracted from the face image for error correction coding and introduces an adaptive random masking scheme to align the bit error rate with error correction requirements, thus mitigating the noise gap between the biometric measurements and ECC error tolerance.

Secondly, a supervised learning-based fifth generation (5G) new radio (NR) low-density parity-check (LDPC) neural decoding scheme, namely Neural-MS decoder, is integrated into \textit{WiFaKey} to enhance the error tolerance of the instrument system. Neural-MS decoder enabled us to achieve a more robust error correction performance with fewer iterations, thereby increasing the reliability and security of the instrument system. 

Our research is pioneering in proposing a random masking scheme that adaptively reduces measurement noise in bio-cryptosystems. Additionally, we are the first to integrate a supervised learning-based 5G NR LDPC Min-Sum neural decoding scheme into a bio-cryptosystem on unconstrained face datasets.
Consequently, we tackled the performance degradation and noise gap problems from two directions by reducing the noise level in biometric measurements and improving the neural decoder of the LDPC. The proposed method can address the challenge and ultimately achieve a successful solution effectively. The main contributions of this paper are as follows:

\begin{itemize}
\item We propose a new fuzzy commitment BC system, named \textit{WiFaKey}. The proposed fuzzy commitment system can handle unconstrained facial biometric measurements, which is more practical for real-world scenarios. 
\item A pipeline called \textit{AdaMTrans} is proposed for transforming features into binary bits suitable for LDPC coding. \textit{AdaMTrans} utilizes quantization and binarization techniques, along with a random bit masking scheme, to effectively address the challenge of biometric intra-class variation while preserving LDPC error correction capabilities with minimal impact on accuracy.
\item A supervised learning-based 5G NR LDPC Min-Sum neural decoding scheme on the binary channel, namely Neural-MS decoder, is proposed for the fuzzy commitment system. The Neural-MS decoder improves error correction performance with less iteration than non-learning decoders.
\end{itemize}

Our experiment simulates practical use cases with independent subjects (no identity overlaps the training and testing datasets) and fully validates the performance benefits of the proposed scheme.
We employed four widely used face feature extractors (Arcface \cite{deng2018arcface}, FaceViT \cite{zhong2021facevit}, Adaface \cite{kim2022adaface}, and Magface \cite{meng2021magface}) and examined its performance across six large unconstrained datasets (LFW \cite{huang-2008-LFW}, CALFW \cite{zheng2017crosscalfw}, CPLFW \cite{zheng2018crosscplfw}, CFPFP \cite{sengupta2016frontalcfpfp}, CFPFF \cite{sengupta2016frontalcfpfp}, and AgeDB-30 \cite{moschoglou2017agedb}) and two constrained datasets (CMU-PIE\cite{sim2002cmu} and color FERET\cite{phillips2000feret}). Comparative analysis with existing state-of-the-art key generation schemes revealed a significant improvement in the performance of \textit{WiFaKey}.

\section{Related Work}
Biometric cryptosystem schemes are developed to address the challenges associated with the use of biometric measurements directly as cryptographic keys. In this section, we first reviewed existing works on ECCs-enabled fuzzy commitment in biometric instruments, encompassing various modalities not limited to face recognition. The impact of ECCs on key retrieval performance is discussed subsequently.  

\subsection{ECCs enabled fuzzy commitment}
\textbf{Hadamard and Reed-Solomon ECC}: An early application of fuzzy commitment to iris biometrics was demonstrated by Hao et al. using a combination of two ECCs, Hadamard and Reed--Solomon \cite{hao-TC-irisfuzzycommit-2006}. For implementing the (64, 7) Hadamard ECC, which can correct at least 15 random bit errors and outputs a 7-bit word, the 2048-bit iris template was divided into 32 blocks of 64 bits each. The Reed--Solomon code (32, 20) was used to wipe out errors at the block level (burst), which can correct up to 6 incorrect 7-bit words. To obtain a 140-bit key, the 7-bit words used 32 input words to decode 20 output words. 
The authors of \cite{rathgeb2010adaptive} integrated a fuzzy commitment scheme into the iris recognition system and proposed a method to enhance accuracy. They employed Hadamard and Reed-Solomon codes for error correction and improved performance by optimizing the arrangement of iris codes based on global error distribution analysis.

\noindent\textbf{Reed-Muller ECCs}: The key recovery rate was found to drop drastically in \cite{hao-TC-irisfuzzycommit-2006} when tested on a challenging dataset. Bringer et al.\cite{bringer2007optimal} proposed an iterative soft decoding ECC as a combination of two Reed-Muller ECCs, (64, 7) and (32, 6), which significantly enhanced the accuracy of the fuzzy commitment scheme.

\noindent\textbf{BCH Codes}:
In \cite{chen2007biometric}, BCH codes were employed to construct a fuzzy commitment scheme for face-ordered feature vectors.
\cite{nandakumar2010fingerprint} applied a fuzzy commitment scheme with turbo codes for securing fingerprint minutiae patterns.
\cite{yang2018securing} utilized a fuzzy commitment scheme based on BCH codes to bind finger-vein templates and secret keys. The resulting helper data was stored on a smart card. In \cite{wu2023multi}, a palmprint fuzzy commitment system based on deep hashing codes generated by a deep hashing network was proposed, and the system is designed on top of BCH.

\noindent\textbf{LDPC Codes}: \cite{li2012effective} introduced a binary fixed-length feature generation strategy based on minute triplets for fingerprint fuzzy commitment. This work explored using LDPC codes (BCH and Reed–Solomon are also explored).
The authors of \cite{chen2019face} devised a face template protection method employing multi-label learning and LDPC codes. This method involves hashing random binary sequences to create a protected template encoded with an LDPC encoder during training to yield diverse binary codes. The face features of each user are subsequently mapped to a diverse binary code using deep multi-label learning. Finally, the LDPC decoder eliminates noise caused by intra-variations from the CNN output.
\cite{dong2021biocancrypto} incorporates LDPC coding into a bio-cryptosystem scheme, utilizing it to extract stable cryptographic keys from finger templates. 

\subsection{Impact of ECCs to FC and motivations of WiFaKey}
ECCs play a crucial role in handling variations in biometric measurements, and their error correction capability must be sufficiently robust to distinguish between intra-class and inter-class variations. However, limitations still exist, as discussed in Section \ref{sec.intro}, which hinders the development of biometric key generation.

Firstly, they require binary biometric vectors of the same size as the employed codewords, necessitating the conversion of biometric features into binary vectors (e.g., \cite{chen2007biometric,li2012effective,chen2019face,wu2023multi}). This conversion can potentially lead to a loss of discriminability. In unconstrained settings, the variability within intra-class and inter-class can be even more pronounced, underscoring the importance of robust error correction mechanisms \cite{sellahewa2010image,rai2022robust}. For instance, in face recognition, datasets such as FRGC \cite{phillips2005overview}, FERET \cite{phillips2000feret}, PIE \cite{sim2002cmu}, and Extended Yale B \cite{georghiades2001fewYaleb} are commonly used in works like \cite{chen2019face, rathgeb2022deep,chen2007biometric}. As a result, fuzzy commitment schemes for unconstrained settings remain relatively underexplored.

Secondly, linear codes such as BCH, Hadamard, RS, LDPC, and their combinations have been extensively studied in the abovementioned literature. However, the error-correcting codes utilized in existing fuzzy commitment schemes for face recognition have not been optimized to their full potential (e.g., non-learning LDPC in \cite{li2012effective,chen2019face,dong2021biocancrypto}). Their error correction capability must be sufficiently robust to differentiate between intra-class and inter-class variations \cite{noto2011analysis}. Nevertheless, the noise gap, as discussed in \ref{sec.intro}, often proves to be substantial. Leveraging known statistics about the differences in the considered biometrics can still prove invaluable in accommodating the high variability inherent in unconstrained scenarios.

In our work, WiFaKey leverages the proposed AdamTrans and Neural-MS decoder to address the limitations above in existing works. We will provide detailed information in subsequent sections.

\section{Proposed Method}

\begin{figure*}[t!]
 \centering
 \includegraphics[width=0.99\linewidth,trim=0cm 6.5cm 10cm 0cm, clip]{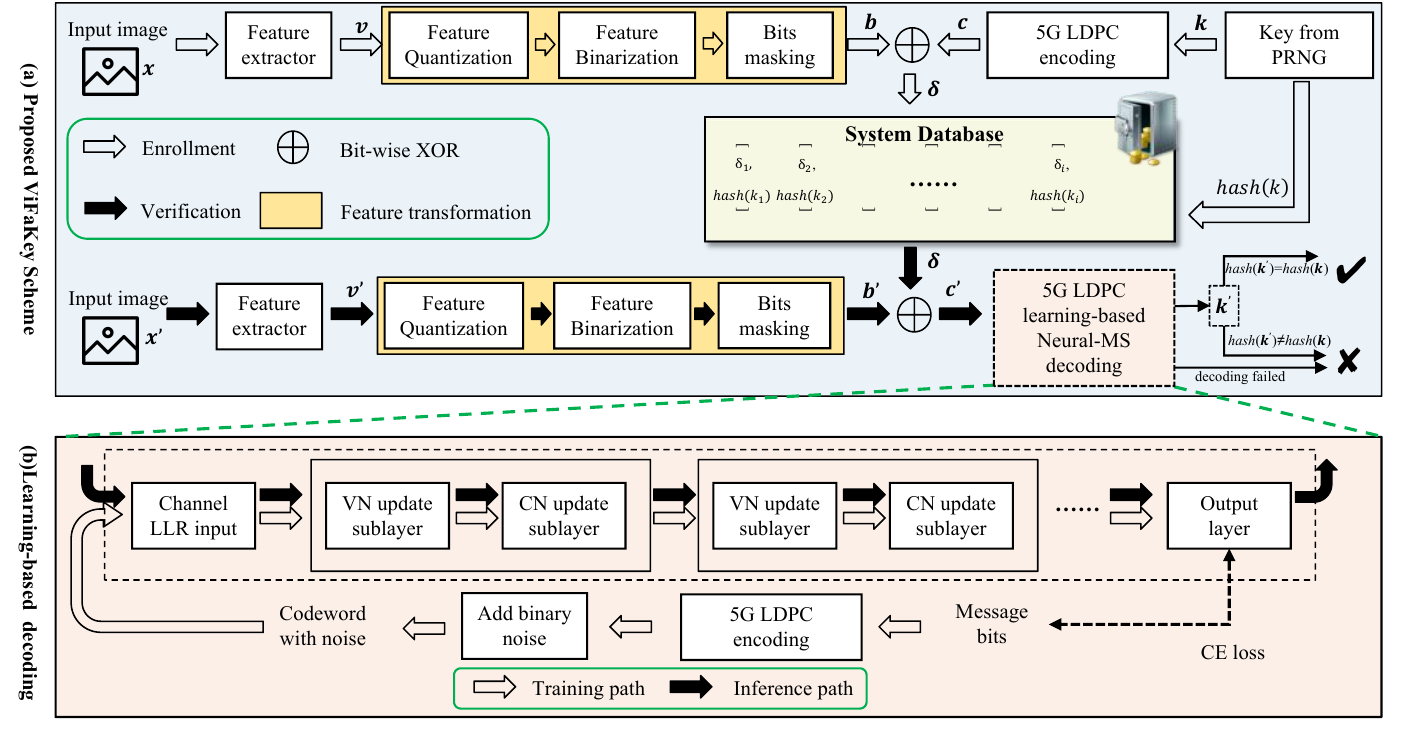} 
 \caption{\textbf{Overview of the \textit{WiFaKey}.}
Our proposed feature transformation pipeline, i.e., the AdaMTrans, transforms features generated from extractors into binary templates. The commitment is generated and stored in the database; a neural min-sum decoder (the lower block) is trained and serves as the decoder to retrieve the key from the de-committed codewords.
\label{Figure.Overview}}
 \vspace{-0.5cm}
\end{figure*}

\subsection{Overview}
An overview of the proposed fuzzy commitment is shown in Fig.~\ref{Figure.Overview}.
At the enrollment stage of a BC scheme, given an input image $\mathbf{x} \in R^{H*W*L}$ where $H$ and $W$ represent the height and width dimensions of $\mathbf{x}$ and $L$ represents the number of channels, highly discriminative features can be extracted using popular pre-trained face recognition models by $\mathbf{v} = f(\mathbf{x}) \in R^{512}$, where $f(\cdot)$ is the feature extractor. Then the features are transformed into binary instances by $\mathbf{b}=g(\mathbf{v}) \in \{0,1\}^n$, where $g(\cdot)$ is the AdaMTrans consisting of feature quantization, binarization, and random masking.
Next, a set of $n$-bit codewords $C$ is generated from a randomly generated key $\mathbf{k}$ using the LDPC coding scheme, and the commitment is computed and stored as \{${hash(\mathbf{k}),\mathbf{\delta}}$\} where $\mathbf{\delta} = \mathbf{b}~\mathrm{XOR}~\mathbf{c}$, $\mathbf{c}$ is a codeword $\mathbf{c} \in C$, and $hash(\cdot)$ is a one-way hashing such as SHA3~\cite{dworkin2015sha}.

At the authentication and key retrieval stage, given a query image $\mathbf{x\prime} \in R^{H*W*L}$ from the same person, a binary instance $\mathbf{b\prime}$ can be obtained following the same pipeline, i.e., $\mathbf{b\prime}=g(f(\mathbf{x\prime}))$.
The stored $\delta$ can be de-committed to recover the codeword by $\mathbf{c}^\prime =\mathbf{b}^{\prime}~\mathrm{XOR}~\mathbf{\delta}$.
The codeword $\mathbf{c}^\prime$ can be restored as $\mathbf{c}$ by the EEC system if the distance between $\mathbf{b^\prime}$ and $\mathbf{b}$ is within the ECC's error correction capability.

The above task can be optimized along three dimensions, focusing on measurement accuracy and precision: 1) minimize the intra-class variations and maximize the inter-class variations of the biometric features; 2) optimize the transformation function $g(\cdot)$ to preserve the discriminative information of the biometric features after binarization; 3) optimize the error correction to achieve the highest error correction capability for intra-class and lower capability for inter-class, ensuring reliable and accurate measurements for identity authentication.
The first dimension has been well-explored, and tremendous feature extraction models have been proposed in the literature.
However, the second dimension remains under-explored except for a few works such as \cite{rathgeb2022deep,drozdowski2018benchmarking}.
These works mainly focus on transforming floating feature vectors into binary vectors, while the gap between bit error rates and ECC capability is overlooked.
As for the optimization of the ECC, to our best knowledge, there is no existing work using a learning technique to optimize ECC for the bio-cryptosystem. This presents an opportunity for further research to enhance the precision and reliability of biometric measurements in cryptographic applications.

To bridge the gap between the bit error rates and ECC capability, we have designed a feature transformation function and a learning-based neural-MS decoder on top of the widely used facial feature extractor. The feature transformation and neural decoder details are discussed in later subsections.

\subsection{Feature transformation\label{sec.adamtrans}}
\begin{figure}[t!]
 \centering
\includegraphics[width=0.9\linewidth, clip]{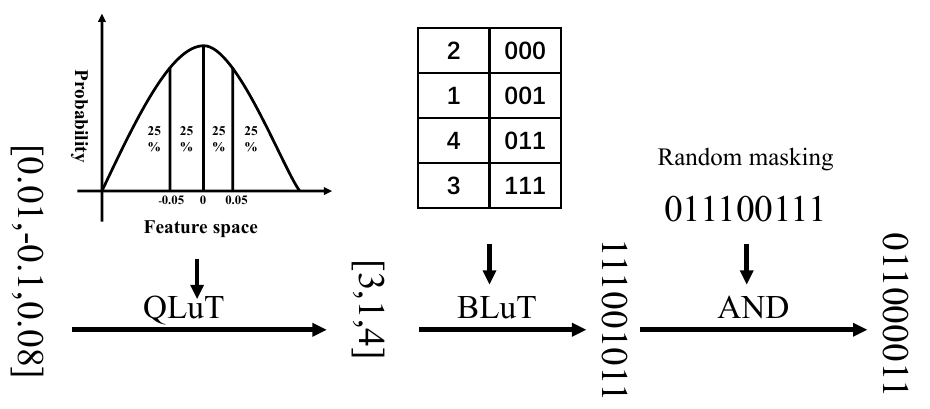} 
 \caption{\textbf{Diagram of the AdaMTrans.}
\label{Figure::feat_bin}}
 \vspace{-0.5cm}
\end{figure}

The feature transformation by AdaMTrans aims to achieve several measurement-focused has several goals: 1) features generated by pre-trained face models are usually fixed-length and real-valued vectors that must be converted into binary strings suitable for ECC; 2) to preserve the discrimination capability and accuracy in the binary domain; 3) given a specific LDPC code scheme, bridge the gap between bit error rates from the same user's binary features and the error correction capabilities of the LDPC.

In AdaMTrans, since comprehensive studies have already been conducted for the first and second goals in \cite{rathgeb2022deep,drozdowski2018benchmarking}, we adopt the feature quantization and binarization approaches mentioned therein, i.e., the quantization-based on equally probable intervals and the linearly separable subcode binarization (LSSC) \cite{lim2012novel} to address the first two goals.
However, the transformation adopted by \cite{rathgeb2022deep,drozdowski2018benchmarking} does not consider the requirements of the ECC, and the intra-class errors of the binary features may be beyond the tolerance of the LDPC (see Section \ref{section.masking}).
Therefore, carefully designed random masking bits are applied through bitwise AND operation with the binary features to align with the ECC correction capability.

\textbf{Feature quantization and binarization}: The feature quantization and binarization can be represented as two lookup tables (see Fig.~\ref{Figure::feat_bin}).
Accordingly, the conversion of the $\mathbf{v}$ to integer $\mathbf{z}$ is accomplished with the following equation:
\begin{equation}
\mathbf{z}=QLuT(\mathbf{v}) \in\{1,q\}^{512},
\end{equation}
Where $QLuT$ refers to the quantization schemes based on equally probable intervals, as suggested by \cite{rathgeb2022deep,drozdowski2018benchmarking}.
Specifically, the feature elements' probability densities are calculated first.
The feature space of each feature element is then divided into $q$ integer-labeled equally probable intervals based on its obtained probability density.
Then, each feature vector element is mapped to an integer corresponding to the interval it belongs to (see an example in Fig.~\ref{Figure::feat_bin}).
Then, the quantized feature vector $\mathrm{z}$ is mapped to a binary feature vector $\mathbf{preb} \in\{0,1\}^{512*m}$ using the second lookup table:
\begin{equation}
\label{eq.blut}
\mathbf{preb}=BLuT_{\varphi}(\mathbf{z}) \in\{0,1\}^{512*m},
\end{equation}
where $BLuT$ is defined as the index-permuted LSSC \cite{lim2012novel} similar to \cite{rathgeb2022deep}, and $\varphi$ denotes the permutation seed; $m=q-1$ is the bit length of the binary entries in $BLuT$ (see an example in Fig.~\ref{Figure::feat_bin}).
Note that the L1 norm of the distance between two binary values generated by LSSC is equivalent to the distance between the two quantized values.

\textbf{Random masking}: On the other hand, we propose a random masking strategy to decrease the intra-class Hamming distance to bridge the gap between the bit error rate from the intra-class binary features and the error correction capability.
Specifically, the masking bits string is generated as a random binary vector $\mathbf{r}$ with $\kappa$ percent of the bits being 0.
Next, each binary vector is bitwise AND with the mask $\mathbf{r}$ to generate the masked binary vector:
\begin{equation}
\mathbf{b}= \mathbf{preb} ~\text{AND}~ \mathbf{r} ,
\end{equation}
where $\mathbf{r}=sgn(\mathbf{u}-\kappa)$, $\mathbf{u} \in \mathcal{U}(0,1)^{512*m}$.
The distance \textbf{D} between any two samples after masking can be computed as follows:
\begin{equation}
\begin{aligned}
\mathbf{D} &= ||( \mathbf{b}~\text{AND}~\mathbf{r} )~\text{XOR}~(\mathbf{b}~\text{AND}~\mathbf{r}) ||_1\\
 & = ||\mathbf{b}~\text{XOR}~\mathbf{b} ||_1 - 0.5 \times ||\mathbf{r}||_1,
\end{aligned}
\end{equation}
therefore, the intra-class and inter-class Hamming distance can be mediated based on different masks $\mathbf{r}$.
A proper mask $\mathbf{r}$ should lead to a smaller intra-class distance within the error correction capability but a large inter-class distance beyond the error correction capability.
To address the latter issue, we iteratively search for the best $\kappa$ which can lead to 95\% of the inter-class distance after the masking is larger than a given threshold $\tau$ (see Section \ref{section.masking}), i.e.,
\begin{equation}
\label{eq.tau}
P( 
||\mathbf{preb_i}~\text{XOR}~\mathbf{preb_j} ||_1 - 0.5\times||\mathbf{r}||_1
 \tau) =0.95,
\end{equation}
where $\mathbf{preb_i}$ and $\mathbf{preb_j}$ are from different users.
Note that the $\mathbf{r}$ is determined based on a ten-fold cross-validation protocol on each dataset, and $\mathbf{r}$ is application-specific in our settings and can be exposed to the public. In practice, $\mathbf{r}$ can also be user-specific, and using a user-specific $\mathbf{r}$ would improve the performance.

On the other hand, raising the $\tau $ parameters can result in a zero false match rate (FMR), but this comes with a lower geniue match rate (GMR) or utility. It's essential to recognize that there is a compromise between the GMR and the FMR, and attaining a 0 FMR may not always be practical in real-world scenarios.
\subsection{Learning-based LDPC neural decoding}
LDPC is a linear error correcting code \cite{gallager1962low}, which can nearly reach the Shannon channel capacity and decode in linear time in the data length based on the iterative belief propagation (BP) algorithm.
LDPC can be described with a sparse parity-check matrix $\mathbf{H}$, which contains very few ones.
The ones in each row from $\mathbf{H}$ specify which bits are associated in the corresponding parity equation.
$\mathbf{H}$ can be used to create the generator matrix $\mathbf{G}$, which encodes a message $\mathbf{a}$ with $\mathbf{aG}~mod~2$ based on matrix multiplication in modulo two arithmetic.

LDPC can also be represented as a sparse Tanner graph \cite{tanner1981recursive}.
A Tanner graph has two sets of nodes: variable nodes (VNs) and check nodes (CNs).
Each row of the parity-check matrix (also known as the parity-check equations) represents a CN, and each bit in the codeword stands for a VN.
In a parity-check equation, the corresponding VN and CN have an edge connected if a bit is involved.

BP decoding algorithms have been widely studied, a.k.a. sum-product (SP) decoding algorithms.
These iterative decoding algorithms operate on the Tanner graph by passing the bit log-likelihood ratio (LLR) on the graph.
Given a noisy received signal vector $\mathbf{y}$ corresponding to a transmitted codeword $\mathbf{x}$, the LLR of the $v$-th bit in $\mathbf{x}$ is defined as
\begin{equation}
\ell_{v}=\ln \frac{\operatorname{Pr}\left(y_{v} \mid x_{v}=0\right)}{\operatorname{Pr}\left(y_{v} \mid x_{v}=1\right)},
\end{equation}
where $\operatorname{Pr}\left(y_{v} \mid x_{v}\right)$ denotes the transition probability from $x_{v}$ to $y_{v}$.
Iterative message exchanges between VNs and CNs commonly describe the SP decoding of LDPC codes.
During iteration $i=1,\cdots, I$, the check-node messages $\mathrm{CN}$ and variable-node $\mathrm{VN}$ message are updated as follow: 
\begin{equation}
\ell_{v \rightarrow c}^{(i)}=\ell_{v}+\sum_{c^{\prime} \in \mathcal{N}(v) \backslash c} \ell_{c^{\prime} \rightarrow v}^{(i-1)},
\end{equation}
\begin{equation}
\ell_{c \rightarrow v}^{(i)}=2 \tanh ^{-1}\left(\prod_{v^{\prime} \in \mathcal{N}(c) \backslash v} \tanh \left(\frac{\ell_{v^{\prime} \rightarrow c}^{(i)}}{2}\right)\right),\label{eq.sptanh}
\end{equation}
where $\ell_{v \rightarrow c}^{(i)}$ and $\ell_{c \rightarrow v}^{(i)}$ denote the message passed along $VN-to-CN$ and $CN-to-VN$ at iteration $i$, respectively.
$\mathcal{N}(v)$ ($\mathcal{N}(c)$) represents the neighboring node set of $v$ ($c$) consisting of CNs (VNs) which are adjacent to the $\mathrm{VN} v$ ($\mathrm{CN} c$); $\mathcal{N}(v) \backslash c$ ($\mathcal{N}(c) \backslash v$) denotes the neighboring node-set except the $\mathrm{CN}$ $c$ ($\mathrm{VN} v$).
Note that $\ell_{c \rightarrow v}^{(0)}$ is initialized to 0.
Note that after $I$ iterations, $v-to-c$ updating is computed differently:
\begin{equation}
s_{v}=\ell_{v}+\sum_{c^{\prime} \in \mathcal{N}(v)} \ell_{c^{\prime} \rightarrow v}^{(i)};
\end{equation}
The estimated code bits can be generated by
\begin{equation}
\hat{x}_{v}=\frac{1-\operatorname{sgn}\left(s_{v}\right)}{2},
\end{equation}
where $\operatorname{sgn}(\cdot)$ represents the signum function.

Hyperbolic tangent functions are used by the SP decoder discussed above.
To alleviate the computational overhead, the min-sum (MS) algorithm approximates Equation (\ref{eq.sptanh}) using
\begin{equation}
\ell_{c \rightarrow v}^{(i)}=\left(\prod_{v^{\prime} \in \mathcal{N}(c) \backslash v} \operatorname{sgn}\left(\ell_{v^{\prime} \rightarrow c}^{(i)}\right)\right) \times \min _{v^{\prime} \in \mathcal{N}(c) \backslash v}\left|\ell_{v^{\prime} \rightarrow c}^{(i)}\right|.
\end{equation}

The MS decoding algorithm has a much lower computational complexity at the cost of slightly lower performance.
To preserve the performance, the normalized min-sum (NMS) has been proposed to add a normalization scaling factor $\alpha$: 
\begin{equation}
\begin{aligned}
\ell_{c \rightarrow v}^{(i)}=&\alpha \times \left(\prod_{v^{\prime} \in \mathcal{N}(c) \backslash v} \operatorname{sgn}\left(\ell_{v^{\prime} \rightarrow c}^{(i)}\right)\right) \times \\
& \max \left\{\min _{v^{\prime} \in \mathcal{N}(c) \backslash v}\left|\ell_{v^{\prime} \rightarrow c}^{(i)}\right|, 0\right\},\label{eq.NMS}
\end{aligned}
\end{equation}
while offset min-sum (OMS) has been proposed to add an offset correction term $\beta$:
\begin{equation}
\begin{aligned}
\ell_{c \rightarrow v}^{(i)}=& \left(\prod_{v^{\prime} \in \mathcal{N}(c) \backslash v} \operatorname{sgn}\left(\ell_{v^{\prime} \rightarrow c}^{(i)}\right)\right) \times \\
& \max \left\{\min _{v^{\prime} \in \mathcal{N}(c) \backslash v}\left|\ell_{v^{\prime} \rightarrow c}^{(i)}\right|-\beta, 0\right\}.
\label{eq.oms}
\end{aligned}
\end{equation}

The normalizing and offset factors listed above can be considered the weights and biases assigned to the graph's edges.
Therefore, constructing a neural network to learn the optimal normalizing and offset parameters automatically is a straightforward strategy.
Several examples of this have been proposed recently.
For example, an SP decoder is trained in \cite{nachmani2016learning} to learn the weight parameters and an OMS decoder is trained in \cite{lugosch2017neural} to learn the offset parameters.
Our proposed novel fuzzy commitment scheme is adopted from \cite{dai2021learning}.
\cite{dai2021learning} proposes a neural-MS decoder that considers both the normalizing factors and offset factors for 5G NR LDPC, defined by
\begin{equation}
\begin{aligned}
\ell_{c \rightarrow v}^{(i)}=& \left(\prod_{v^{\prime} \in \mathcal{N}(c) \backslash v} \operatorname{sgn}\left(\ell_{v^{\prime} \rightarrow c}^{(i)}\right)\right) \times \\
& \max \left\{\alpha \times \min _{v^{\prime} \in \mathcal{N}(c) \backslash v}\left|\ell_{v^{\prime} \rightarrow c}^{(i)}\right|-\beta, 0\right\}.
\label{eq.neuralms}
\end{aligned}
\end{equation}
To learn the weight and offset parameters for LDPC codes, \cite{dai2021learning} proposed a sparse MS neural decoder with a partially connected structure that resembles a trellis during iterative decoding.

In particular, the neural network input layer is a vector with size $N$, where $N$ is the length of a code block (i.e., the number of variable nodes in the Tanner graph).
Except for the final layer (i.e., all the hidden layers), the remaining layers in the trellis are all of the size $E$, where $E$ represents the number of edges in the Tanner graph.
The hidden layer $i$ corresponds to the $i$-th iteration in the MS decoding process.
Two sublayers, $i_{v}$ and $i_{c}$, which stand for the updates of $\mathrm{VN}$ and $\mathrm{CN}$, respectively, are included in each hidden layer.
There are $N$ neurons in the output layer.
In the hidden layer $i$, sublayer $i_{v}$ outputs the message sent from the associated $\mathrm{VN}$ to $\mathrm{CN}$ along the edge, and sublayer $i_{c}$ outputs the message sent from the associated $\mathrm{CN}$ to $\mathrm{VN}$ along the edge (refer to \cite{dai2021learning} for more details).

The neural network is trained greedily in an iteration-by-iteration manner to avoid the gradient vanishing issue \cite{dai2021learning}, i.e., once a layer is trained, its weights and biases are fixed for subsequent training iterations. The training starts from a one-layer network and gradually grows to a multi-layer network. 

We adopt the base code BG2 defined in the 5G standard \cite{5Gldpc}, which has 42 rows and 52 columns in the parity-check matrix $\mathbf{H}$ with 197 nonzero elements, and the baseline code rate of the BG2 codes is $0.2$. However, the decoder is revised to be suitable for biometric measurements, making it distinct from \cite{dai2021learning} where: \textbf{1) the original neural-MS decoder is designed for the AWGN channel.
In contrast, our decoder is designed to work on the binary symmetric channel (BSC) for the binary operation in fuzzy commitment.
2) we adopt the lifting factor $z=10$, therefore the code length is $N=52\times10$ and the message length is $K=10\times10$, which is appropriate for our features length. } 

The binary features $\mathbf{b}$ are divided into $m$ (from (\ref{eq.blut})) blocks with a block size of 512.
Each block is padded with eight zeros to fit the code length 520, and $m$ codewords will be generated in the encoding process.
In consequence, a key length of $100\times m$ can be achieved.
In the implementation, the $m$ subkeys are retrieved from $m$ codewords and then concatenated to form the final key \textbf{k}. We follow the training details as discussed above to train the decoder greedily in an iteration-by-iteration manner.

\section{Experiments and Results}
\subsection{Unconstrained datasets}
We mainly report the recognition performance on six mainstream benchmarks, including LFW \cite{huang-2008-LFW}, CALFW \cite{zheng2017crosscalfw}, CPLFW \cite{zheng2018crosscplfw}, CFPFP \cite{sengupta2016frontalcfpfp}, CFPFF \cite{sengupta2016frontalcfpfp}, and AgeDB-30 \cite{moschoglou2017agedb}. These are all unconstrained face datasets and the experimental results based on unconstrained face datasets are beneficial for emphasizing the feasibility of the proposed method in practical applications.

\begin{itemize}
 \item LFW database, a well-known unconstrained face dataset, has 13,233 face images from 5,749 identities.
LFW was used to build the Cross-Age LFW (CALFW) and Cross-Pose LFW (CPLFW) databases, which focus on the challenges to face recognition of adversarial robustness and similar-looking faces.
 \item Celebrities in Frontal-Profile (CFP) in the Wild contains faces from 500 celebrities in frontal and profile views.
This dataset presents two verification protocols: one comparing only frontal faces (CFP FF) and the other comparing frontal and profile faces (CFP FP).
 \item AgeDB is a cross-age database that was manually collected. It contains 16,488 images from 568 celebrities of different ages.
AgeDB-30 is a sub-dataset of this, containing identities with an age difference of 30 years, and is the most challenging one.
\end{itemize}
CFPFF and CFPFP contain 3,500 genuine and 3,500 impostor comparisons, and LFW, CALFW, CPLFW, and AgeDB30 have 3,000 genuine and 3,000 impostor comparisons.
In addition, we have conducted a comparison with existing works on two widely used face datasets: color FERET (colorFERET) \cite{phillips2000feret}, and CMU-PIE \cite{sim2002cmu}. colorFERET consists of face images captured in a semi-controlled environment with 13 different poses, making it a semi-constrained dataset. On the other hand, CMU-PIE also belongs to the semi-constrained category and includes 67 subjects with 42 illumination variations in the database.

\subsection{Feature extractors}
Four popular pre-trained models are adopted in our experiments, including Arcface \cite{deng2018arcface} (resnet100 \cite{he2016deep} trained on ms1mv3\cite{deng2019lightweight}), Face ViT \cite{zhong2021facevit} (P8S8 trained on ms1m-retinaface \cite{deng2019lightweight} with cosface loss function), Adaface \cite{kim2022adaface} (ir101 on webface12m \cite{zhu2021webface260m}), and Magface \cite{meng2021magface} (ir100 trained on MS1MV2 \cite{deng2019lightweight}). The details of the adopted pre-trained models are summarised in the supplementary Table \ref{sup.pretrain}. It is worth highlighting that the employed neural-network-based feature extractors are tested on a disjoint dataset of subjects, which falls naturally into open-set and subject-independent recognition settings. 

\subsection{On the error correction capability \label{section.capbility}}

Due to the inherent complexity associated with determining the decoding radius of LDPC codes—a challenge often equated to the NP-hard problem of identifying the minimum distance of a linear ECC—the performance of iterative decoders, which are employed for the decoding of LDPC codes, is typically evaluated numerically through comprehensive Monte Carlo simulations. Monte Carlo simulations statistically approximate the LDPC code's performance metrics, such as the bit-error-rate (BER), under specific conditions like a given signal-to-noise ratio (SNR). The results can be averaged by conducting many simulations under varied conditions to yield a reliable estimate of the LDPC code's error correction capability.

In this study, the performance of the SP, MS, and neural-MS decoders is evaluated over a Binary Symmetric Channel (BSC). The BSC experiences different crossover rates, which represent the probabilities of bit-flipping. The methodology adopted involves encoding 10,000 randomly generated data frames (10,000 $\mathbf{k}$) using an LDPC code encoder. After the encoding, varying levels of noise, dictated by the crossover rate, are introduced to the codeword. After introducing noise, the noisy message is decoded using the SP, MS, and neural-MS decoders. The Frame Error Rate (FER), the ratio of received messages containing errors to the total number of transmitted messages, is used in this assessment. Note that a single-bit error within a frame can render the entire frame erroneous, resulting in the FER typically being higher than the BER for a specific code and channel condition. This observation is particularly beneficial for thoroughly evaluating the error correction capacity of the decoders under consideration. Furthermore, it provides the definitive upper limit for decoding error, an essential factor for successful key retrieval.

\begin{figure*}[t!]
 \centering
 \includegraphics[width=0.8\linewidth,trim=0cm 11cm 23cm 0cm, clip]{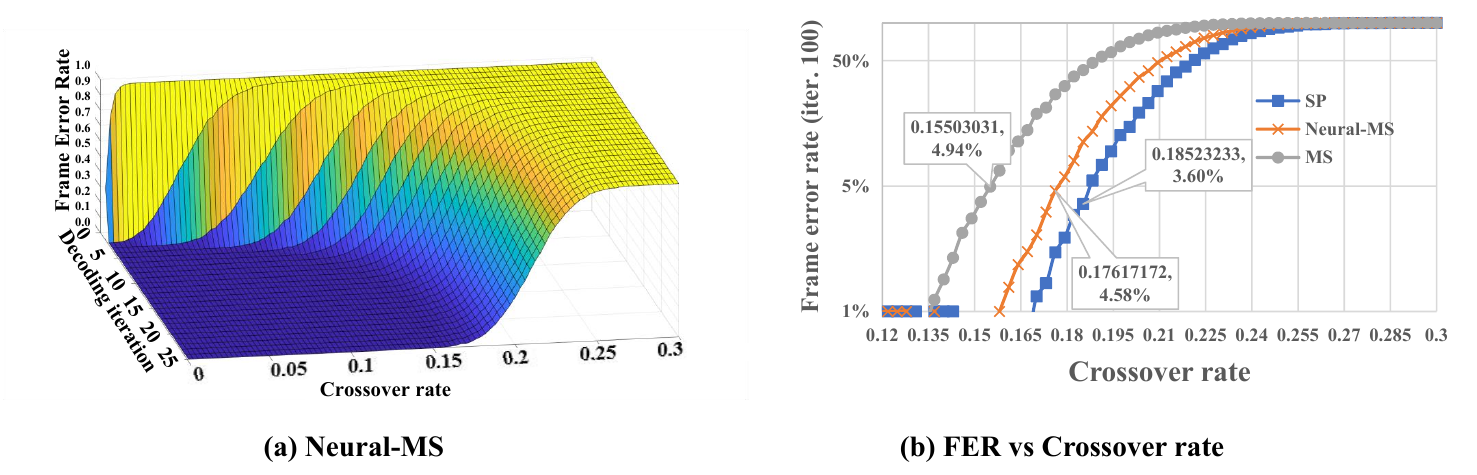} 
 \caption{\textbf{Frame error rate vs.~crossover rate and number of iterations.}
\label{Figure::FERcscrossovervsiter}}
 \vspace{-0.5cm}
\end{figure*}

Figure \ref{Figure::FERcscrossovervsiter} (a) and Figure \ref{Figure::supFERcscrossovervsiter} (a-c) shows the FER under different crossover rates and iterations. Figure \ref{Figure::supFERcscrossovervsiter} is included in the appendix.
Fig.~\ref{Figure::FERcscrossovervsiter} (b) shows the FER vs.~different crossover rates after 100 iterations.
The results show that neural-MS outperforms the handcrafted MS decoder by a large margin yet is comparable to the SP decoder.
SP is computationally unfriendly due to the (inverse) hyperbolic tangent function used in its decoding process, whereas neural-MS strikes a balance between computational overhead and accuracy.

 
As seen from Fig.~\ref{Figure::FERcscrossovervsiter} (b), to achieve a FER less than 5\%, MS works favorably up to a 15.5\% crossover rate, Neural-MS can tolerate a 17.62\%, while SP can work well up to an 18.52\%.
Therefore, it is fair to conclude that an intra-class Hamming distance within 18.52\% can be suitable for LDPC coding.
However, the intra-class distance is usually larger than this FER (see Table \ref{table.binfeat}); this is why we introduce the AdaMTrans, elaborated in Section \ref{sec.adamtrans}.

\subsection{Ablation studies\label{section.masking}}
\textbf{Binary features recognition performance on LSSC intervals $q$}:
We first evaluate the recognition performance of the generated binary features under various LSSC intervals $q$.
The comparison scores between pairs of deep face representations are obtained for deep facial features by estimating their cosine distance.
The recognition performance is evaluated in terms of accuracy, which represents the probability of the fuzzy commitment BC correctly identifying the identity,
Decidability ($d'$) \cite{daugman2000biometric} is also used to indicate the separability of mated and non-mated\footnote{Mated corresponds to intra-user, Non-mated corresponds to inter-user, these terms are used interchangeably.} comparison scores. The $d'$ is defined as
\begin{equation}
 d^{\prime}=\frac{\left|\mu_m-\mu_{nm}\right|}{\sqrt{\frac{1}{2}\left(\sigma_m^2+\sigma_{nm}^2\right)}},
\end{equation}
where $\mu_m$ and $\mu_{nm}$ are the means of the mated and the non-mated comparison trials score distributions, and $\sigma_m$ and $\sigma_{nm}$ are their standard deviations, respectively. The larger the decidability values, the better the ability to distinguish between mated and non-mated comparison trial scores. 

As shown in Table \ref{table.binfeat}, consistent with \cite{rathgeb2022deep,drozdowski2018benchmarking}, the LSSC can achieve comparable performance to the deep features in terms of accuracy.
Dividing the feature space into larger intervals generally leads to better performance.
However, the performance reaches a plateau quickly using more than eight intervals.

\begin{table*}
\centering
\caption{Ablation study of feature transformation.\label{table.binfeat}}
\resizebox{\linewidth}{!}{%
\begin{tabular}{c|l|cccc|cccc|cccc} 
\hline
 & & accuracy & $d'$& Mated & non-mated & accuracy & $d'$& Mated & non-mated & accuracy & $d'$& Mated & non-mated \\
\cdashline{3-14}
 & & \multicolumn{4}{c|}{agedb\_30} & \multicolumn{4}{c|}{lfw} & \multicolumn{4}{c}{cfp\_fp} \\ 
 \hline
\multirow{5}{*}{Magface} & Deep features & $98.28\pm0.81$& 4.81 & $23.09\pm6.38$& $48.86\pm4.07$& $99.80\pm0.24$& 8.17 & $13.46\pm4.92$& $49.19\pm3.75$& $98.40\pm0.56$& 4.43 & $23.99\pm6.84$& $49.01\pm4.12$\\
 & LSSC-2 (sign) & $97.92\pm0.86$& 4.26 & $31.60\pm4.98$& $49.27\pm3.11$& $99.75\pm0.25$& 6.56 & $23.47\pm4.76$& $49.42\pm2.94$& $97.71\pm0.71$& 3.78 & $32.18\pm5.37$& $49.32\pm3.51$\\
 & LSSC-4 & $98.18\pm0.87$& 4.46 & $26.92\pm3.93$& $41.06\pm2.17$& $99.77\pm0.23$& 6.86 & $20.14\pm3.82$& $41.13\pm2.02$& $97.93\pm0.73$& 3.88 & $27.42\pm4.24$& $41.09\pm2.61$\\
 & LSSC-8 & $98.20\pm0.87$& 4.49 & $24.42\pm3.47$& $36.89\pm1.84$& $99.80\pm0.24$& 6.90 & $18.31\pm3.43$& $36.97\pm1.70$& $98.27\pm0.59$& 3.90 & $24.87\pm3.74$& $36.92\pm2.26$\\
 & LSSC-16 & $98.18\pm0.84$& 4.50 & $23.03\pm3.25$& $34.70\pm1.70$& $99.77\pm0.26$& 6.92 & $17.29\pm3.21$& $34.77\pm1.56$& $98.20\pm0.58$& 3.91 & $23.47\pm3.50$& $34.74\pm2.10$\\ 
\hline
\multirow{5}{*}{Adaface} & Deep features & $98.00\pm0.75$& 4.76 & $26.09\pm6.13$& $49.32\pm3.17$& $99.82\pm0.23$& 8.57 & $14.88\pm5.08$& $49.90\pm2.77$& $99.26\pm0.46$& 5.36 & $23.86\pm5.88$& $49.75\pm3.47$\\
 & LSSC-2 (sign) & $97.62\pm0.75$& 4.12 & $33.91\pm4.68$& $49.55\pm2.64$& $99.75\pm0.23$& 6.57 & $24.93\pm4.81$& $49.94\pm2.41$& $98.83\pm0.56$& 4.39 & $32.34\pm4.65$& $49.81\pm3.18$\\
 & LSSC-4 & $97.88\pm0.72$& 4.32 & $28.93\pm3.66$& $41.38\pm1.79$& $99.78\pm0.21$& 6.91 & $21.36\pm3.84$& $41.65\pm1.59$& $99.10\pm0.61$& 4.55 & $27.58\pm3.64$& $41.55\pm2.36$\\
 & LSSC-8 & $98.02\pm0.75$& 4.38 & $26.22\pm3.23$& $37.24\pm1.50$& $99.83\pm0.20$& 6.93 & $19.42\pm3.44$& $37.42\pm1.30$& $99.14\pm0.53$& 4.59 & $24.99\pm3.21$& $37.34\pm2.04$\\
 & LSSC-16 & $98.00\pm0.85$& 4.39 & $24.73\pm3.02$& $35.02\pm1.37$& $99.83\pm0.20$& 6.93 & $18.32\pm3.23$& $35.19\pm1.19$& $99.16\pm0.48$& 4.60 & $23.56\pm3.00$& $35.12\pm1.90$\\ 
\hline
\multirow{5}{*}{FaceViT} & Deep features & $97.37\pm0.94$& 4.10 & $20.11\pm4.79$& $36.59\pm3.06$& $99.67\pm0.28$& 6.63 & $12.45\pm4.12$& $36.41\pm3.03$& $95.86\pm0.90$& 3.63 & $20.20\pm5.08$& $35.52\pm3.12$\\
 & LSSC-2 (sign) & $96.83\pm0.94$& 3.81 & $34.54\pm4.80$& $49.20\pm2.56$& $99.62\pm0.37$& 5.81 & $26.60\pm5.04$& $49.50\pm2.39$& $95.74\pm1.13$& 3.37 & $35.25\pm5.45$& $49.46\pm2.40$\\
 & LSSC-4 & $97.73\pm0.66$& 4.03 & $27.84\pm3.47$& $38.75\pm1.62$& $99.65\pm0.35$& 5.95 & $21.58\pm3.80$& $38.81\pm1.52$& $96.11\pm0.82$& 3.57 & $28.17\pm3.78$& $38.53\pm1.58$\\
 & LSSC-8 & $97.45\pm0.69$& 4.05 & $24.69\pm3.00$& $34.15\pm1.38$& $99.62\pm0.33$& 5.96 & $19.19\pm3.32$& $34.19\pm1.29$& $96.34\pm0.71$& 3.60 & $24.94\pm3.24$& $33.88\pm1.36$\\
 & LSSC-16 & $97.52\pm0.76$& 4.05 & $23.11\pm2.79$& $31.91\pm1.28$& $99.62\pm0.32$& 5.97 & $17.97\pm3.08$& $31.93\pm1.21$& $96.24\pm0.73$& 3.62 & $23.30\pm3.00$& $31.62\pm1.26$\\ 
\hline
\multirow{5}{*}{Arcface} & Deep features & $98.38\pm0.72$& 5.04 & $23.10\pm6.36$& $49.58\pm3.84$& $99.82\pm0.27$& 7.97 & $14.71\pm5.11$& $49.88\pm3.58$& $98.70\pm0.55$& 4.72 & $24.42\pm6.41$& $49.73\pm4.05$\\
 & LSSC-2 (sign) & $97.88\pm0.89$& 4.42 & $31.60\pm4.99$& $49.73\pm2.95$& $99.68\pm0.34$& 6.40 & $24.65\pm4.80$& $49.94\pm2.86$& $98.01\pm0.67$& 3.99 & $32.67\pm4.98$& $49.83\pm3.50$\\
 & LSSC-4 & $97.97\pm0.81$& 4.65 & $27.00\pm3.92$& $41.53\pm2.03$& $99.75\pm0.30$& 6.64 & $21.18\pm3.90$& $41.64\pm1.95$& $98.26\pm0.54$& 4.10 & $27.87\pm3.95$& $41.56\pm2.57$\\
 & LSSC-8 & $98.23\pm0.70$& 4.69 & $24.54\pm3.46$& $37.38\pm1.73$& $99.82\pm0.27$& 6.70 & $19.28\pm3.48$& $37.49\pm1.64$& $98.54\pm0.49$& 4.13 & $25.28\pm3.49$& $37.37\pm2.23$\\
 & LSSC-16 & $98.27\pm0.78$& 4.69 & $23.15\pm3.24$& $35.14\pm1.59$& $99.83\pm0.24$& 6.71 & $18.22\pm3.26$& $35.24\pm1.50$& $98.49\pm0.56$& 4.13 & $23.85\pm3.26$& $35.14\pm2.07$\\
\hline
\end{tabular}
}
\end{table*}

\begin{figure*}[t!]
 \centering
 \includegraphics[width=0.99\linewidth,trim=0cm 11cm 5.5cm 0cm, clip]{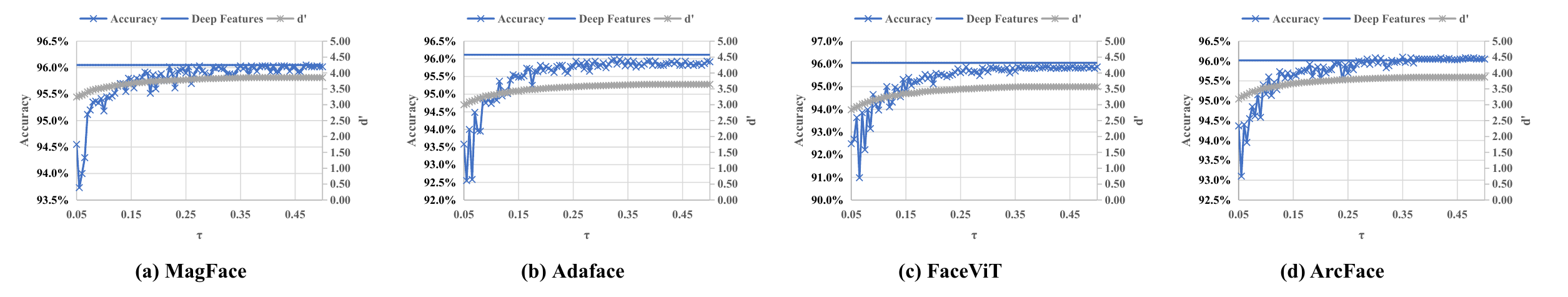} 
\vspace{-0.6cm}
 \caption{\textbf{Ablation study on $\tau$ concerning accuracy on transformed binary features (CALFW).}\label{Figure::accvstau}}
  \vspace{-0.5cm}
\end{figure*}

\noindent\textbf{$\tau$ on different feature extractors}:
The parameter $\tau$ in Eq.(\ref{eq.tau}) is used to control the generation of the masking bits, and we test the accuracy of the $\mathbf{b}$ after masking with $\tau = [0.05,0.5]$.
As shown in Fig.~\ref{Figure::accvstau}, the accuracy improves according to the increase in $\tau$.
It can also be observed that the performance is primarily preserved compared to the original deep features.
For example, the bit masking operation on features from Magface and Arcface can achieve an accuracy similar to their deep features counterparts. Applying the masking operation to Adaface and FaceViT only leads to a 0.3\% and 0.5\% drop in accuracy.

\begin{figure*}[t!]
 \centering
 \includegraphics[width=0.99\linewidth,trim=0cm 11cm 0cm 0cm, clip]{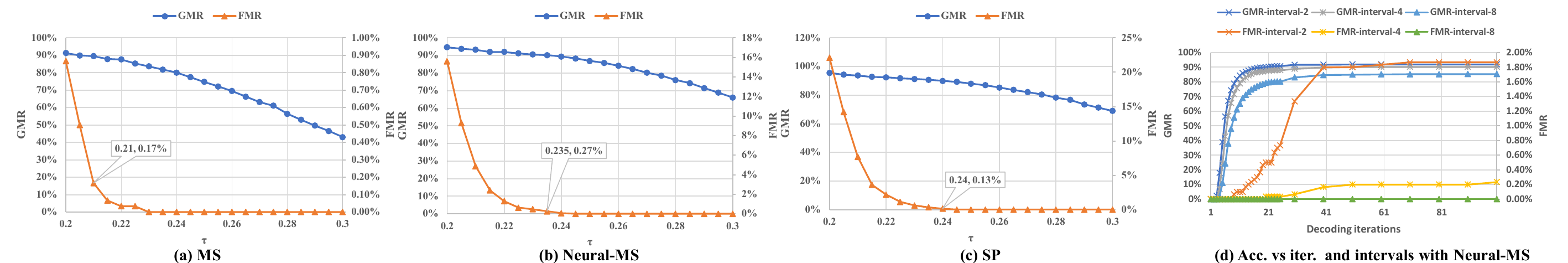} 
\vspace{-0.6cm}
\caption{\textbf{Ablation study on $\tau$ with respect to key retrieval accuracy} (CALFW, 100 iterations).
\label{Figure::tauvsGMRablation}}
\vspace{-0.1cm}
\end{figure*}

\noindent\textbf{$\tau$ on different decoders}:
We must decide which $\tau$ is the best for each decoder (MS, NeuralMS, and SP).
Features of CALFW extracted from Magface are adopted as the testing dataset, and the key retrieval performance in genuine match rate (GMR, i.e., 1 minus FNMR) and false match rate (FMR) is evaluated.
The result is shown in Fig.~\ref{Figure::tauvsGMRablation} (a-c).

It can be seen that FMR drops to zero as $\tau$ increases, but the GMR will also degrade.
The $\tau$ that drives FMR just above 0.1\% is selected to balance FMR and GMR.
It can be seen that MS can achieve a 0.17\% FMR when $\tau=0.21$, and neural-MS can achieve a 0.27\% FMR at $\tau=0.235$, while SP can earn a 0.13\% FMR at $\tau=0.24$.
Those $\tau$ are adopted for the best parameters for MS, neural-MS, and the SP decoder for all datasets and feature extractors subsequently. Further experiment results are shown in appendix section \ref{section.supmasking}.

\noindent\textbf{Key retrieval performance on different LSSC intervals $q$}:
Based on the best parameter of $\tau$, the key retrieval performance under different numbers of intervals ($q$) in LSSC is evaluated.
The result is shown in Fig.~\ref{Figure::tauvsGMRablation} (d).
It can be seen that neural-MS achieves its highest GMR when $q=2$, yet the FMR deteriorates to around 1.8\%.
Its GMR becomes lowest when $q=8$, with a perfect FMR.
However, a balance between GMR and FMR was achieved when $q=4$, with a GMR close to that for $q=2$ but a very low FMR close to that for $q=8$.
Therefore, we adopted $q=4$ ($m=3$) for the remaining experiments.

\begin{figure*}[t!]
\centering
\includegraphics[width=0.99\linewidth,trim=0.6cm 12.8cm 7.5cm 0.7cm, clip]{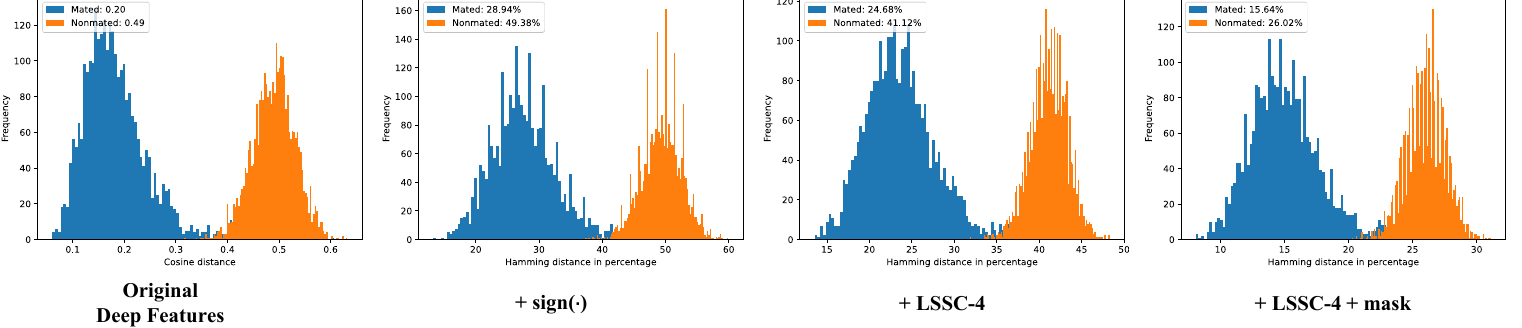} 
\vspace{-0.4cm}
\caption{\textbf{Distribution of distances across different stages} (CALFW, Magface).\label{Figure::dist_process}}
\vspace{-0.1cm}
\end{figure*}

\noindent\textbf{Distributions on different stages}:
Fig.~\ref{Figure::dist_process} shows the distributions, at each stage, of the mated and non-mated Hamming distances.
It can be seen that the feature transformation using LSSC and random masking preserves the separation of mated and non-mated distributions.
Take AgeDB30 as an example: the mated and non-mated distances are 26.92\% and 41.06\%, which has a gap of 14.14\%, corresponding to $d'=4.46$, which is comparable to its deep feature counterpart $d'=4.81$.
We can also observe that the random masking (+ LSSC-4 + mask) can reduce the Hamming distance to comply with the requirements of the LDPC.
As a result, the mated distances can be reduced to 15.64\% and 17.02\% on CALFW and AgeDB30, which is within the error correction capability and suitable for the adopted LDPC.
Meanwhile, the non-mated distance is reduced to 26.02\% and 25.93\% on CALFW and AgeDB30, which is still beyond the error correction capability.

\subsection{Key retrieval performance}
\begin{figure*}[t!]
\centering
\includegraphics[width=0.99\linewidth,trim=2cm 22cm 2cm 0cm, clip]{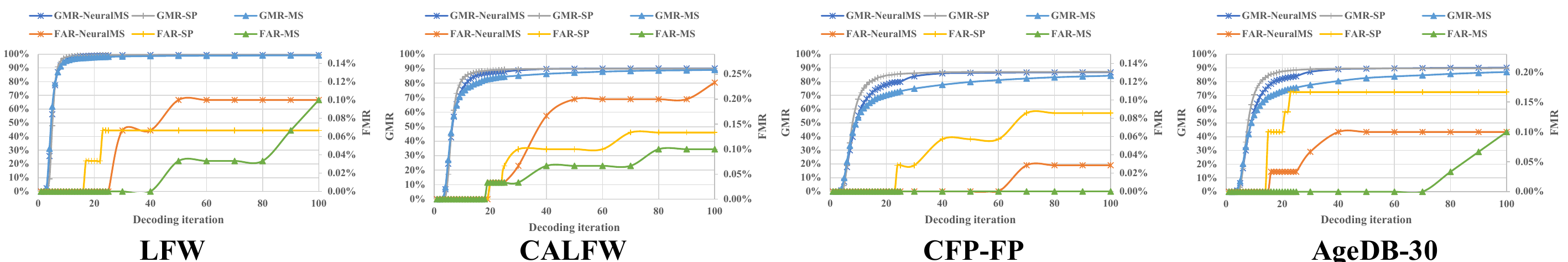} 
\vspace{-0.2cm}
\caption{\textbf{Key retrieval performance in GMR and FMR vs. decoding iterations} (MagFace).
\label{Figure::MagfaceGMR}}
\end{figure*}

This section evaluates the performance under decoding iterations ranging from 1 to 100.
The performance of different decoding strategies is shown in Fig.~\ref{Figure::MagfaceGMR}.

We can observe that: 1) both GMR and FMR increase concerning decoding iteration; 2) when the decoding iteration reaches 20 and above, the GMR and FMR level off; 3) good performance can be achieved on LFW regardless of the feature extractor, while the other datasets, i.e., CALFW, AgeDB30, and CFPFP, can also achieve around 90\% accuracy.
Note that all benchmarks are unconstrained face datasets; hence, it is reasonable to infer that the proposed method is feasible in practice for secret key management; 4) the choice of an appropriate
number of decoding iterations can make the system's accuracy suitable for deployment.
The operator can opt for small iteration numbers to avoid a high false accept rate.
More iterations are preferred if the system is to achieve a higher GMR without considering the FMR.

Detailed quantitative results are summarized in Table \ref{table.keyretrieve}.
As we can see from the results, MS performs the worst, as it is a handcrafted decoding method.
Thanks to the learned weights and offset parameters, neural-MS performs better than the MS decoder and is comparable with the SP decoder.

On the other hand, we can observe that both neural-MS and SP achieve a favorable performance on LFW.
Specifically, NeuralMS can reach 99.17\%(99.5\%), 99.33\%(99.67\%), 99.37\%(99.70\%), 98.77\%(99.27\%) on Magface, Arcface, Adaface, and FaceVit in terms of GMR@0FMR(deep features).
CFP-FF can also achieve a performance similar to that of LFW.

On the other hand, there is a significant performance drop in CPLFW, AgeDB30, and CFPFP.
The $d'$ from Table \ref{table.binfeat} also indicates that AgeDB30 and CFPFP are worse than LFW (4.81, 4.43 vs.~8.17).
Therefore, it is reasonable to observe the performance drop on CPLFW, AgeDB30, and CFPFP.

We also show the GMR and FMR at 100 iterations in Table \ref{table.keyretrieve} (GMR\textbar{}FMR@iter100 column).
Under this setting, the GMR of CPLFW, AgeDB30, and CFPFP can be improved significantly, only slightly compromising on FMR.
For example, Arcface can achieve 76.50\% (16.90\%) at FMR=0.17\% (0FMR) CPLFW, and 92.33\% (86.83\%) at FMR=0.13\%(0FMR).
Similar observations can also be made for the other feature extractors.

\begin{table*}
\centering
\caption{Accuracy of key retrieval performance.\label{table.keyretrieve}}
\resizebox{\linewidth}{!}{%
\begin{tabular}{c|ccccc|ccc|ccccc|ccc} 
\hline
\multirow{2}{*}{Dataset} & \begin{tabular}[c]{@{}c@{}}Deep \\Feat.~\end{tabular} & LSSC-4 & MS & SP & Neural-MS & MS~ & SP & Neural-MS & \begin{tabular}[c]{@{}c@{}}Deep \\Feat.~\end{tabular} & LSSC-4 & MS & SP & Neural-MS & MS~ & SP & Neural-MS \\
 & \multicolumn{5}{c|}{GMR@FMR0} & \multicolumn{3}{c|}{GMR\textbar{}FMR@iter100} & \multicolumn{5}{c|}{GMR@FMR0} & \multicolumn{3}{c}{GMR\textbar{}FMR@iter100} \\ 
\hline
 & \multicolumn{8}{c|}{Magface} & \multicolumn{8}{c}{Arcface} \\
LFW & 99.50 & 99.60 & 98.67 & 99.03 & 99.17 & 99.17\textbar{}0.10 & 99.47\textbar{}0.07 & 99.50\textbar{}0.10 & 99.67 & 99.63 & 98.07 & 99.33 & 99.37 & 99.20\textbar{}0.07 & 99.43\textbar{}0.07 & 99.37\textbar{}0.00 \\
CALFW & 91.10 & 90.87 & 82.40 & 88.80 & 87.13 & 89.23\textbar{}0.10 & 90.03\textbar{}0.13 & 90.23\textbar{}0.23 & 91.53 & 91.13 & 81.97 & 85.27 & 85.30 & 89.20\textbar{}0.23 & 90.43\textbar{}0.20 & 90.40\textbar{}0.13 \\
CPLFW & 58.80 & 60.80 & 43.37 & 53.23 & 64.73 & 72.97\textbar{}0.17 & 75.30\textbar{}0.10 & 75.90\textbar{}0.17 & 56.13 & 58.10 & 9.57 & 18.60 & 16.90 & 74.30\textbar{}0.07 & 77.53\textbar{}0.23 & 76.50\textbar{}0.17 \\
AgedDB30 & 94.30 & 91.43 & 84.73 & 83.13 & 76.73 & 87.10\textbar{}0.10 & 89.87\textbar{}0.17 & 90.20\textbar{}0.10 & 96.13 & 95.27 & 85.27 & 89.73 & 86.83 & 90.40\textbar{}0.10 & 92.63\textbar{}0.07 & 92.33\textbar{}0.13 \\
CFP-FP & 99.43 & 99.43 & 84.34 & 85.31 & 86.60 & 84.34\textbar{}0.00 & 87.17\textbar{}0.09 & 86.83\textbar{}0.03 & 96.77 & 97.03 & 82.26 & 88.09 & 82.37 & 85.94\textbar{}0.06 & 90.29\textbar{}0.09 & 89.69\textbar{}0.06 \\
CFP-FF & 99.43 & 95.06 & 94.91 & 98.29 & 98.31 & 99.03\textbar{}0.14 & 99.46\textbar{}0.20 & 99.34\textbar{}0.17 & 99.63 & 99.57 & 97.00 & 97.83 & 98.54 & 98.83\textbar{}0.09 & 99.20\textbar{}0.09 & 99.26\textbar{}0.09 \\
Average & 90.43 & 89.53 & 81.40 & 84.63 & 85.45 & - & - & - & 89.98 & 90.12 & 75.69 & 79.81 & 78.22 & - & - & - \\ 
\hline
 & \multicolumn{8}{c|}{Adaface} & \multicolumn{8}{c}{FaceViT} \\
LFW & 99.70 & 99.67 & 98.67 & 99.37 & 98.57 & 99.03\textbar{}0.07 & 99.43\textbar{}0.03 & 99.30\textbar{}0.07 & 99.27 & 99.60 & 97.93 & 98.60 & 98.77 & 98.43\textbar{}0.07 & 98.77\textbar{}0.03 & 98.77\textbar{}0.00 \\
CALFW & 90.90 & 91.10 & 84.40 & 76.03 & 82.23 & 86.63\textbar{}0.10 & 88.63\textbar{}0.13 & 88.03\textbar{}0.13 & 86.50 & 90.90 & 76.13 & 85.23 & 84.53 & 85.40\textbar{}0.17 & 86.63\textbar{}0.17 & 86.50\textbar{}0.03 \\
CPLFW & 76.37 & 81.10 & 61.10 & 75.50 & 69.30 & 79.03\textbar{}0.13 & 80.97\textbar{}0.13 & 81.17\textbar{}0.07 & 6.70 & 22.93 & 26.57 & 8.33 & 33.83 & 67.43\textbar{}0.13 & 70.13\textbar{}0.13 & 69.23\textbar{}0.07 \\
AgedDB30 & 95.43 & 94.63 & 82.80 & 77.03 & 72.60 & 82.80\textbar{}0.00 & 87.10\textbar{}0.03 & 86.20\textbar{}0.13 & 88.57 & 92.93 & 52.73 & 74.70 & 77.10 & 79.63\textbar{}0.07 & 82.67\textbar{}0.13 & 82.40\textbar{}0.07 \\
CFP-FP & 98.29 & 98.06 & 87.69 & 92.97 & 90.57 & 90.54\textbar{}0.06 & 92.97\textbar{}0.00 & 93.11\textbar{}0.03 & 65.77 & 91.46 & 66.94 & 73.31 & 66.69 & 76.09\textbar{}0.11 & 77.94\textbar{}0.09 & 77.49\textbar{}0.09 \\
CFP-FF & 99.83 & 99.71 & 97.23 & 96.91 & 97.94 & 98.89\textbar{}0.09 & 99.09\textbar{}0.11 & 98.91\textbar{}0.11 & 99.40 & 99.40 & 95.17 & 97.69 & 97.71 & 97.89\textbar{}0.11 & 97.97\textbar{}0.03 & 97.71\textbar{}0.00 \\
Average & 93.42 & 94.05 & 85.31& 86.30 & 85.20 & - & - & - & 74.37 & 82.87 & 69.25 & 72.98 & 76.44 & - & - & - \\
\hline
\end{tabular}
}
\end{table*}

\subsection{Comparison with other SOTA methods}
\begin{table*}
\centering
\caption{Comparison with relevant face-based secure systems.\label{tab.sotacompare}}
\resizebox{\linewidth}{!}{%
\begin{tabular}{llcccccc} 
\hline
\multirow{2}{*}{Reference(year)} & \multicolumn{1}{c}{\multirow{2}{*}{Category}} & \multicolumn{4}{c}{Performance} & \multirow{2}{*}{Security} & \multirow{2}{*}{unlinkability} \\ 
\cline{3-6}
 & \multicolumn{1}{c}{} & LFW & CFP & ColorFERET & CMUPIE & & \\ 
\hline
Kumar et al. \cite{kumar2018face} (2018) & CNN & & & 86.92\%@0.01\%FMR & 91.91\%@0.01\%FMR & & \\
Chen et al. \cite{chen2019face} (2019) & CNN & & & & 98.8@0.1\%FMR & & \\
Mai et al. \cite{mai2020secureface} (2021) & \begin{tabular}[c]{@{}l@{}}feature \\transformation\end{tabular} & & 85.36\%@0.1\%FMR & 98.55\%@0.1\%FMR & 99.00\% & 56 bits & \begin{tabular}[c]{@{}c@{}}0.0145\\(cfp)\end{tabular} \\
Boddeti et al. \cite{boddeti2018secure} (2018) & \begin{tabular}[c]{@{}l@{}}homomorphic \\encryption\end{tabular} & 96.74\%@0.1\%FMR & & & & 128 to 192 bits & \\
Dong et al. \cite{DONG2020ACMTOMM} (2021) & fuzzy vault & Rank1=99.9\% & & & & 37bits & \\
Rathgeb et al.\cite{rathgeb2022deep} (2022) & fuzzy vault & & & \begin{tabular}[c]{@{}c@{}}1\%FNMR@0.01\%FMR\\(FERET)\end{tabular} & & \textasciitilde{}32 bits & \\
Zhang et al. \cite{zhang2021facial} (2021) & fuzzy extractor & 30\%@2.1e-7FMR & & & & 45.5bits & \\
Gilkalaye et al. \cite{gilkalaye_euclidean_distance_2019} (2019) & \begin{tabular}[c]{@{}l@{}}fuzzy \\commitment\end{tabular} & 36.94\%@0.05\% & & & & & \\
Ours & \begin{tabular}[c]{@{}l@{}}fuzzy \\commitment\end{tabular} & 99.17\%@0\%FMR & 98.31\%@0\%FMR & 99.53\%@0\%FMR & 99.53\%@0.1\%FMR & \begin{tabular}[c]{@{}c@{}}91bits\\(cfp)\end{tabular} & \begin{tabular}[c]{@{}c@{}}0.0063\\(cfp)\end{tabular} \\
\hline
\end{tabular}
}
\vspace{-0.5cm}
\end{table*}

In this section, following the suggestion from \cite{hahn2022biometric}, we compared our results with state-of-the-art secure face templates evaluated on constrained face benchmarking datasets, including CMU-PIE and Color FERET, and on unconstrained datasets, including LFW and CFP. We used the CFP-FF dataset, which was adopted by \cite{mai2020secureface}. We used 7140 facial images of 68 subjects for the CMU-PIE dataset. In the case of ColorFERET, we excluded the profile left and right samples and used the remaining samples for evaluation. In the case of LFW, we used the standard protocol of 6000 pairs for assessment.

Table \ref{tab.sotacompare} shows that the proposed fuzzy commitment scheme outperforms existing works on performance and security on both constrained and unconstrained datasets. Compared to key generation schemes, \cite{gilkalaye_euclidean_distance_2019} achieves a 36.94\% FNMR at 0.05\% FMR, while the method proposed in \cite{DONG2020ACMTOMM} achieves a top-1 accuracy of 99.9\%. Additionally, the fuzzy vault scheme from \cite{rathgeb2022deep} can achieve a 1\% FNMR at 0.01\% FMR, and the fuzzy extractor in \cite{zhang2021facial} can reach a 70\% FNMR at an FMR of 2.1e-7. Compared to these key generation counterparts, our proposed method can achieve better results while maintaining a higher level of security (detailed in Section \ref{sec.sec}).


We want to emphasize that the homomorphic encryption approach proposed in \cite{boddeti2018secure} is designed to perform matching in the encryption domain. On the other hand, in the CNN approaches offered in \cite{kumar2018face} and \cite{chen2019face}, each user is assigned a unique binary code that is used only for training the deep CNN during the enrollment phase. Thus, re-training is needed if a new user is added to the system, making them impractical for deployment. Furthermore, the CNN approach proposed in \cite{mai2020secureface} aims to perform matching in the transformed domain, which differs from ours.

We attribute the success of our proposed method to two factors: first, the neural LDPC decoder can tolerate errors; second, the random masking scheme bridges the gap between the error correction capability of ECC and the original noise from the input features. In a nutshell, our proposed method generates a key from the biometric measurements, which allows for a more secure and accurate authentication process.

\section{Security analysis\label{sec.sec}}
The security risks considered in this paper originate from compromising the stored commitment \cite{mai2020secureface,li2012effective}.
Under this assumption, the adversary can try inverting the commitment back to the input face image.
The successfully inverted face image can access the system as the corresponding enrolled subject.
However, directly synthesizing an image based on a brute-force attack is infeasible due to the vast number of combinations of pixel values.
Yet the adversary can invert the commitment to the image using learning-based reconstruction methods such as \cite{dong2022reconstruct,dong2021towards,mai2018reconstruction}.
These methods usually take the deep face features as the input and output reconstructed images.
However, such reconstruction models cannot be learned directly because our proposed commitment depends on the input images $\boldsymbol{x}$ and the subject-specific random keys $\boldsymbol{k}$.
Therefore, the key $\boldsymbol{k}$ should be known first to achieve the reconstruction.
However, the key in our scheme is secured using one-way hashing: the adversary needs to guess the key $\boldsymbol{k}$ by directly guessing $\boldsymbol{k}$, which has a strength of $2^{100*m}$, or guessing the intermediate feature $\boldsymbol{b}$.
Therefore, the security depends on the easier way of guessing $\boldsymbol{k}$ and $\boldsymbol{b}$. We have formalized the following adversary model to provide a more rigorous security analysis.

\subsection{Adversary Model}
Given that fuzzy commitment schemes utilize ECC, it is essential to formally define our adversary model using information-theoretic measures, i.e., entropy, as defined by Dodis et al. \cite{dodis2004fuzzy}. We quantify the adversary advantage in terms of the leftover entropy, which captures the remaining randomness or uncertainty in the biometric feature $\boldsymbol{b}$ after the adversary has successfully decommitted the commitment $\delta$. Since knowledge of $\boldsymbol{b}$ implies knowledge of $\boldsymbol{k}$, this leftover entropy characterizes the randomness of the key that remains given that the adversary can recover the codeword $c'$ and 
find a collision such that $\mathbf{c'} = \mathbf{c}$ for successful decomitment of $\delta$.

Within this model, we focus on binary symmetric channels, for our adversary model in LDPC decoding. This choice is motivated by three key reasons:
\begin{enumerate}
 \item {LDPC codes are highly efficient for BSCs because they can correct a high percentage of bit errors at low error rates, low decoding complexity, robustness against channel noise, and flexibility in design with various code rates. This provides significant advantages over other channel types and helps prevent overestimation of system security.}
 \item{BSCs are a well-established model for communication channels that exhibit noise. They offer transparency when analyzing the impact of noise on the security of fuzzy commitment schemes using information-theoretic measures such as entropy and channel capacity.}
 \item{BSCs are a valuable reference channel for comparing the security of different schemes, serving as a reliable benchmarking tool. Furthermore, a decoder that performs well on a BSC is likely to perform well in more complex and realistic channels, making BSCs crucial for evaluating the security of a system.}
\end{enumerate}

Based on the above reasons, it can be inferred that the adversary model utilized in our scheme is much more relaxed than the generic security model. This is because we impose more restrictive conditions on the noise in the biometric measurements, specifically, that the noise follows an independent and identically distributed (i.i.d.) distribution. Furthermore, as the security of our scheme assumed the successful decoding and decommitment of $\delta$, it is commonly believed that the decoding problem will be more complex and challenging in a generic channel. 

\subsection{Leftover Entropy Estimation}
To compute the leftover entropy of the biometric feature, it is essential to quantify the randomness or uncertainty of the biometric measurements. This is done by measuring entropy, which is key in designing and evaluating secure biometric systems. To estimate the entropy of a biometric system, we employ a widely used approach that involves comparing the interclass histogram with a Binomial distribution having the same mean $E_{HD}$ and variance $V^2_{HD}$, as commonly done for iris and face biometrics \cite{daugman2000biometric, mai2020secureface}. The similarity between the observed distribution and the Binomial distribution is used to assume that the entropy of the observed distribution is the same as that of the Binomial distribution. Although this technique is necessarily heuristic, it is effective in practice for noisy biometric measurements where interclass distribution is often well-behaved as normally distributed. As depicted in Figure \ref{Figure::dist_process}, it can be seen that the interclass (non-mated) distribution closely resembles the bell shape of a binomial distribution with a particular mean value.

We utilize the entropy of the Binomial distribution as an approximation for the entropy of the observed distribution. The entropy of the Binomial distribution can be calculated using the following equations for degrees of freedom (DOF):
\begin{equation}
 DOF=\frac{E_{H D}\left(1-E_{H D}\right)}{V_{H D}^{2}},
\end{equation}
where $E_{H D}$ and $V_{H D}$ are the mean and standard deviation of the impostor Hamming distance generated from the intermediate feature $\boldsymbol{b}$. 

It is common for the entropy estimate to vary across different biometric measurement sets, as it captures useful information about the underlying biological process and noise. However, simply measuring the degrees of freedom does not provide sufficient information about the effect of noise on the system's security. A more thorough approach is to compute the entropy $H$, given a particular error rate, which is expected as $E_{HD}$ follows::
\begin{equation} H=DOF*{((-E_{HD})\log(E_{HD})-(1-E_{HD})\log(1-E_{HD}))}.
\end{equation}
The above equation describes the entropy of the biometric feature with i.i.d noise.

\begin{figure*}[h]
\centering
\includegraphics[width=0.99\linewidth,trim=0cm 19.5cm 5cm 0cm, clip]{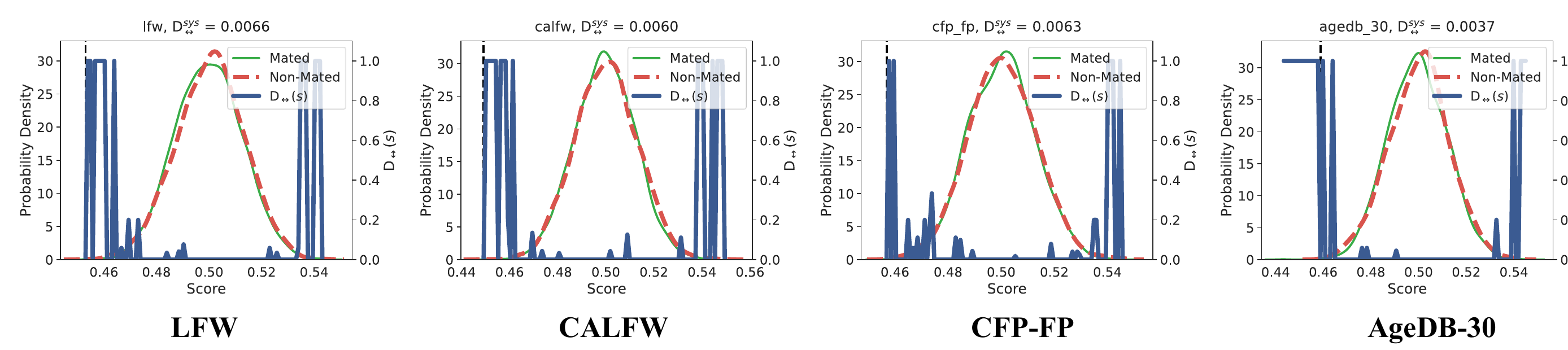} 
\caption{\textbf{Unlinkability evaluation} (Magface).
\label{Figure::unlinkmagface}}
\vspace{-0.5cm}
\end{figure*}
\textbf{Derivation of Lower Bound Leftover Entropy}: As Gallager \cite{gallager1962low} proves, when the degree of CN and VN of an LDPC code increases, the code rate vs. minimum distance trade-off 
of the code approaches the Gilbert-Varshamov (GV) bound. Viewing that individual biometric feature $\boldsymbol{b}$ as a unique codeword, this implies that one could construct a linear ECC for a fuzzy commitment scheme (in worst-case) using $\boldsymbol{b}$ that is capable of achieving minimum
\begin{equation}
s=\frac{2^{H}}{\sum_{i=0}^{d-1}\left(\begin{array}{l}
H \\
i
\end{array}\right)},
\end{equation}
with $d$ and $s$ denoting a code's minimum distance and size (i.e., number of codewords). 

Making use of the fact that for $d \leq \frac{H}{2} : \sum_{i=0}^{d} {H \choose i} \leq 2^{Hh_2(d/H)}$, where $h_2(.)$ is the binary entropy function. We can derive a tighter lower bound for the fuzzy commitment scheme as follows:
\begin{equation}
s={\frac{2^{H}}{\sum_{i=0}^{d}\left(\begin{array}{l}
H \\
i
\end{array}\right)}}\geq{\frac{2^{H}}{\sum_{i=0}^{d-1}\left(\begin{array}{l}
H \\
i
\end{array}\right)}}
\end{equation}
\begin{equation}
\Rightarrow{s=\frac{2^H}{\sum_{i=0}^{d}{{H}\choose{i}}}>{\frac{2^H}{2^{Hh_2(d/H)}}}=2^{H(1-h_2(d/H))}},
\end{equation}
It is important to note that explicitly constructing a linear code to achieve the above lower bound on the security of a fuzzy commitment scheme requires precise estimation of the feature entropy $H$. However, due to the inherent variability and noise in biometric measurements, accurately estimating $H$ can be difficult in practice. As a result, we rely on the assumption of the existence of such a linear code rather than explicitly constructing it. Despite this limitation, the lower bound derived from the assumed code offers valuable insights into the minimum achievable security of a fuzzy commitment scheme with the best ECC from a theoretical perspective.

Given the above result, it is appropriate to designate $d/H$ as the error rate, equivalent to the crossover probability discussed in Section \ref{section.capbility}, which determines the number of errors in the codeword. For a given arbitrary $H$, The best decoder shall be able to tolerate these errors (e.g., given the sufficient high degree of CN and CN) for an LDPC code; thus, a higher crossover probability signifies that the decoder requires a larger $d$ to tolerate the error while maintaining a desirable FER for reliable key retrieval. This exhibits the notable trade-off between security and the number of tolerable errors in the decoder, derivable from the GV bound, where a stronger security guarantee (higher $s$) necessitates a smaller $d$, and vice versa. 

Besides, it is also important to consider the FER to ensure a reliable key retrieval success rate. As depicted in Figure \ref{Figure::FERcscrossovervsiter}, to achieve a FER of less than 4.58\%, it is necessary to consider $d$ such that the condition $d/H\leq{0.1761}$ is satisfied. The result of the minimum achievable security $(s)$ equivalent to $2^{0.33{H}}$. 

Summing up, the security strength of our proposed scheme can be quantitatively defined as
\begin{equation}
H_{s y s}=\min \lbrace{100\times{m}, s\rbrace}.
\end{equation}
In this equation, $100\times{m}$ represents the security length of the random key within our neural-MS framework.

To obtain a more concrete security claim for our proposed scheme, we examine the scenario depicted in Figure \ref{Figure::FERcscrossovervsiter}, where the achievable FER for the neural-MS decoder is less than 5\%. This suggests that the maximum crossover rate must equal $d/H = {0.1761}$. Given that the codeword error must result from a distance between the biometric feature $\boldsymbol{b}$ and $\boldsymbol{b}$, we can apply the lower bound derived previously to determine the value of $s$ by substituting $d = 274 = 1560 \times 0.1761$. This value corresponds to the maximum permissible feature distance for the neural-MS decoder with an FER below 5\%.

It is crucial to note that we employ a conservative calculation for $d$, based on the inequality ${n \times 0.1761}\geq{H \times 0.1761 }$. This approach prevents the overestimation of the system's security. Table \ref{table.security} shows the detailed security strength of neural-MS on various datasets and feature extractors.
The results suggest that Adaface and FaceViT can achieve 100-bit security on most datasets, while Magface and Arcface achieve lower security than Adaface and FaceViT. Take LFW as an example. The proposed fuzzy commitment can achieve 87 bits, 128 bits, 132 bits, and 93 bits when using Magface, Adaface, FaceViT, and Arcface as the extractor.

\begin{table}[h]
\centering
\caption{The security strength ($H_{sys}$ \textbar{} ($s, 100*m$) ) of neural-MS based fuzzy commitment.\label{table.security} }
\resizebox{\linewidth}{!}{%
\begin{tabular}{c|c|c|c|c} 
\hline
 & Magface & Adaface & FaceViT & Arcface \\ 
\hline
LFW & 87 \textbar{} (87,300) & 128 \textbar{} (128,300) & 132 \textbar{} (132,300) & 93 \textbar{} (93,300) \\
CALFW & 87 \textbar{} (87,300) & 122 \textbar{} (122,300) & 137 \textbar{} (137,300) & 103 \textbar{} (103,300) \\
CPLFW & 105 \textbar{} (105,300) & 147 \textbar{} (147,300) & 143 \textbar{} (143,300) & 112 \textbar{} (112,300) \\
CFP-FP & 99 \textbar{} (99,300) & 141 \textbar{} (141,300) & 144 \textbar{} (144,300) & 108 \textbar{} (108,300) \\
CFP-FF & 91 \textbar{} (91,300) & 131 \textbar{} (131,300) & 121 \textbar{} (121,300) & 101 \textbar{} (101,300) \\
AgeDB30 & 103 \textbar{} (103,300) & 151 \textbar{} (151,300) & 145 \textbar{} (145,300) & 106 \textbar{} (106,300)\\
\hline
\end{tabular}
}
\vspace{-0.5cm}
\end{table}

\subsection{Unlinkability evaluation}
In this subsection, we focus on the evaluation of unlinkability. As we generate a random key for each user per enrollment, unlinkability, and cancellability can be naturally guaranteed.
Once the stored commitment is compromised, a new instance can be generated using a different random key.
Besides, using different keys for different enrollments ensures that no information will be leaked from a cross-matching attack, which attempts to infer any information by matching two commitments from two other applications.

We follow a protocol outlined by \cite{gomez2017general}.
Using the Neural-MS parameters, the experiments were executed with Magface and on the LFW, Agedb, and CFP datasets.
A cross-matching attack can be launched by comparing the binary commitment generated from the same person in different applications (different $\mathbf{k}$).
Under this attack, the adversary is assumed to hold the commitments of the other applications.
The adversary can exploit the matching score distributions of the commitments to learn whether the commitment is from the same person.
Here, we refer to the matching distance of the same subject in different applications as the mated distance.
In contrast, the matching distance from other subjects in various applications is the non-mated distance.

In \cite{gomez2017general}, two distinct measures for linkability are defined, namely a local measure $D_{\leftrightarrow}(s) \in[0,1]$ and a global measure $D_{\leftrightarrow}^{sys}$.
Each specific linkage scores $s$ is evaluated by the $D\leftrightarrow(s)$ on a score-wise level.
The unlinkability of the whole system is assessed by $D_{\leftrightarrow}^{sys} \in[0,1]$, which can be used as a standard for various systems regardless of the score.
$D_{\leftrightarrow}^{sys}=1$ indicates the system is fully linkable for all scores of the mated subjects, while $D_{\leftrightarrow}^{sys}=0$ suggests the system is fully unlinkable for all scores. A detailed explanation is given in Appendix \ref{section.appendixunlink}.

The mated and non-mated scores are computed on the selected dataset based on the official protocols among those commitments, and their distributions are illustrated in Fig.~\ref{Figure::unlinkmagface} and Appendix Fig. \ref{Figure::unlinkall}.
It is proved that our scheme can achieve good unlinkability: $D_{\leftrightarrow}^{sys}$ can earn 0.01 for all datasets. This indicates that even if an adversary can obtain commitments from different applications, they cannot distinguish whether they originate from the same subject or different subjects.

\section{Discussion and Conclusion}
\textbf{Failure cases.} 
We identified several factors contributing to key generation failures from the same individual, including variations in pose, hairstyle, lighting, and makeup, which introduce significant noise into the facial measurements. These factors highlight the challenges of generating robust cryptographic keys in unconstrained settings. Future work will focus on improving the feature extraction and transformation pipeline and data enhancement strategies. 

\textbf{Limitations and future work.} Our work focuses on mitigating performance degradation and addressing the noise gap through the introduction of WiFaKey. However, there is a crucial need for further exploration into quantifying the noise inherent in biometrics, as it significantly impacts key retrieval performance. Ensuring system security, particularly achieving a 0\% False Match Rate, presents an ongoing challenge despite our current average Genuine Match Rate of approximately 8\%. Moving forward, our future efforts will concentrate on developing more effective error correction algorithms to bolster key retrieval performance. Additionally, we aim to explore the integration of facial biometrics with other modalities, such as fingerprint and palmprint, to enhance the robustness of cryptographic key generation.

In conclusion, this paper proposes a new fuzzy commitment scheme, \textit{WiFaKey}, incorporating a deep face feature extractor and a learning-based LDPC decoder, aimed at enhancing biometric measurement accuracy and reliability. Additionally, we introduce AdaMTrans, a feature transformation method designed to convert deep features into binary features suitable for LDPC coding. Our experiments demonstrated that the proposed system outperformed state-of-the-art methods on unconstrained face datasets, highlighting its practical utility for authentication purposes in various applications, including financial services, Internet of Things (IoT) services, and public services.


\bibliographystyle{IEEEtran} 
\bibliography{references}

\begin{IEEEbiography}[{\includegraphics[width=1in,height=1.25in,clip,keepaspectratio]{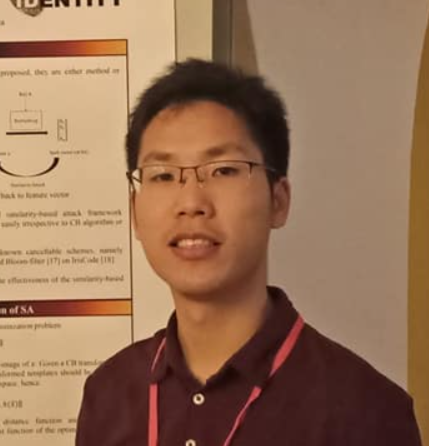}}]{Xingbo Dong}
obtained his B.S. degree from Huazhong Agriculture University, China, in 2014 and his Ph.D. degree from the Faculty of Information Technology, Monash University in 2021. Later, he worked as a Post-Doc at Yonsei University in 2022. Currently, he is a lecturer at Anhui University China. His research interests include biometrics, medical imaging, and image processing.
\end{IEEEbiography}
\vspace{-1.0cm}
\begin{IEEEbiography}
[{\includegraphics[width=1in,height=1.25in,clip,keepaspectratio]{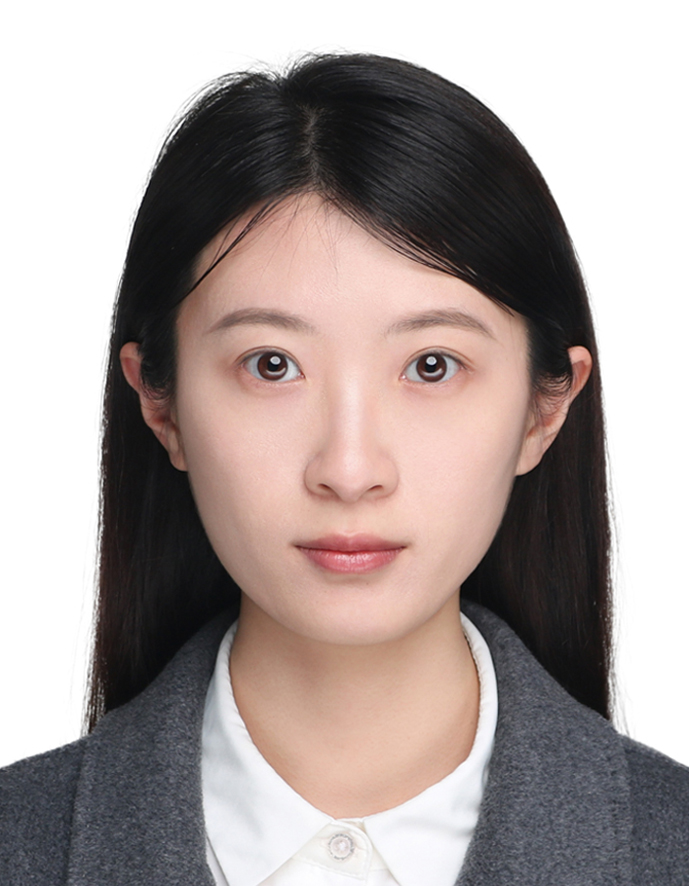}}] 
{\bf{Hui Zhang}} obtained her M.S. degree from Anhui Normal University 
  in 2022 and is currently working toward a PhD degree at Anhui University, China. Her research interests include trustworthy biometrics, biometric template protection, and biometric authentication.
\end{IEEEbiography}
\vspace{-1.0cm}
\begin{IEEEbiography}[{\includegraphics[width=1in,height=1.25in,clip,keepaspectratio]{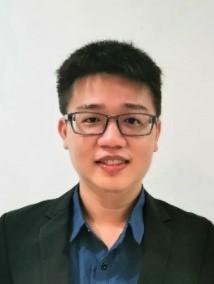}}]{Yen Lung Lai}
obtained his Ph.D. degree from Monash University in 2021 and a BSc in Physics from the University of Tunku Abdul Rahman (UTAR), Malaysia, in 2015. Prior to his current position as a Post-Doc at the School of Artificial Intelligence, Anhui University, China, he worked as an Assistant Professor at University Tunku Abdul Rahman (UTAR) in 2022. His research interests include biometrics security and cryptography.
\end{IEEEbiography}
\vspace{-1.0cm}
\begin{IEEEbiography}[{\includegraphics[width=1in,height=1.25in,clip,keepaspectratio]{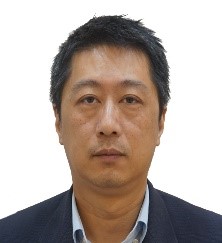}}]{Zhe Jin}
(Member, IEEE) obtained a Ph.D. in Engineering from Universiti Tunku Abdul Rahman Malaysia (UTAR). He is a Professor at the School of Artificial Intelligence, Anhui University, China. His research interests include Biometrics, Pattern Recognition, Computer Vision, and Multimedia Security. He has published over 70 refereed journals and conference articles, including IEEE Trans. IFS, SMC-S, DSC, PR. He was awarded the Marie Skłodowska-Curie Research Exchange Fellowship. He visited the University of Salzburg, Austria, and the University of Sassari, Italy, respectively, as a visiting scholar under the EU Project IDENTITY 690907.
\end{IEEEbiography}
\vspace{-1.0cm}
\begin{IEEEbiography}[{\includegraphics[width=1in,height=1.25in,clip]{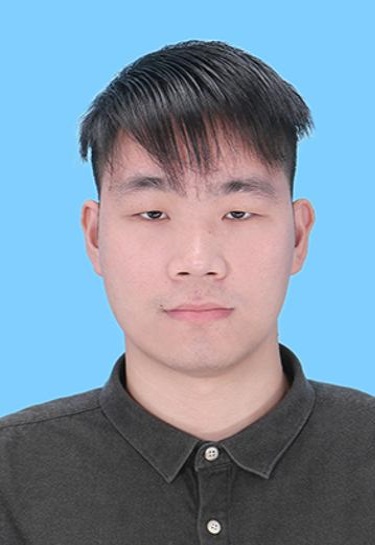}}]{Junduan Huang}
is currently an associate researcher at the School of Artificial Intelligence, South China Normal University, Foshan, China. He received the B.S. degree and M.S. degree from South China Agriculture University, Guangzhou, China, in 2017 and 2020, respectively, and the PhD degree from South China University of Technology, Guangzhou, China, in 2024, respectively. From 2022 to 2023, he was a research intern at Idiap Research Institute, Martigny, Switzerland. His research interests include biometrics, computer vision, audio signal processing, deep learning, and agricultural engineering.
\end{IEEEbiography}
\vspace{-1.0cm}
\begin{IEEEbiography}[{\includegraphics[width=1in,height=1.25in,clip]{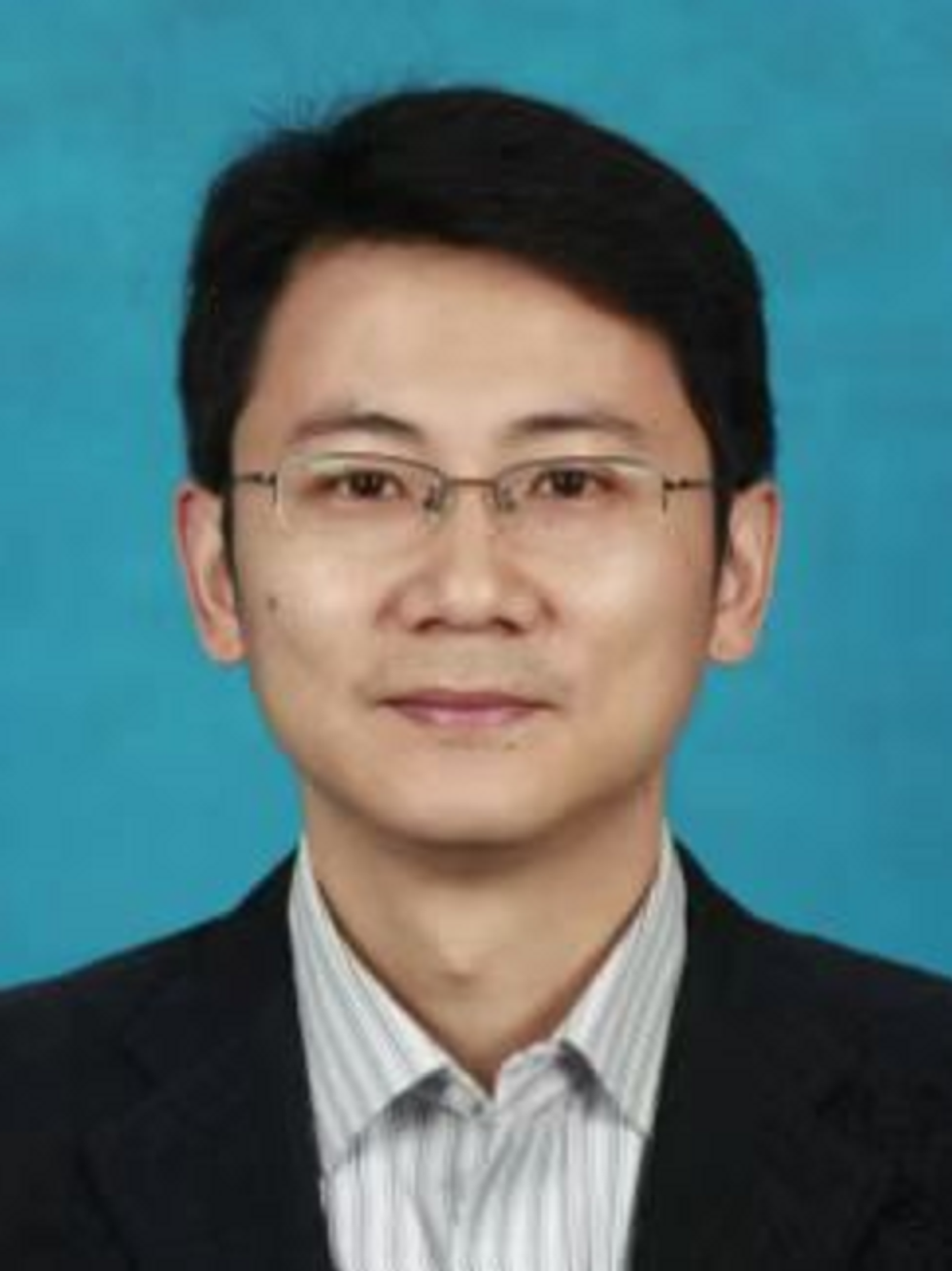}}]{Wenxiong Kang}
received the M.S. degree from Northwestern Polytechnical University, Xi’an, China, in 2003, and the Ph.D. degree from the South China University of Technology, Guangzhou, China, in 2009. He is a Professor at the School of Automation Science and Engineering, South China University of Technology. His research interests include biometrics, image processing, pattern recognition, and computer vision.
\end{IEEEbiography}
\vspace{-1.0cm}
\begin{IEEEbiography}[{\includegraphics[width=1in,height=1.25in,clip,keepaspectratio]{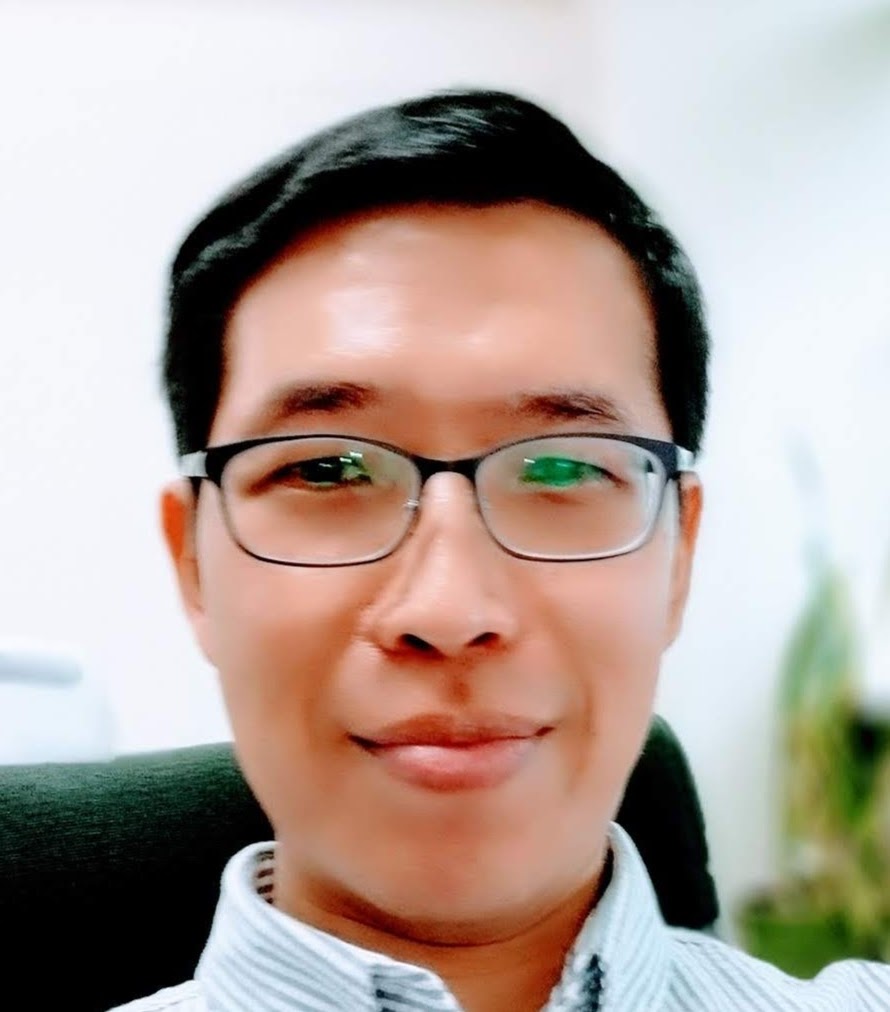}}]{Andrew Beng Jin Teoh}
Andrew Beng Jin Teoh received the B. Eng. (electronic) and Ph.D. degrees from the National University of Malaysia in 1999 and 2003, respectively. He is a full professor in the Electrical and Electronic Engineering Department, College of Engineering of Yonsei University, South Korea. His research, for which he has received funding, focuses on biometric applications and biometric security. His current research interests include machine learning and information security. He has published over 300 international refereed journal papers and conference articles, edited several book chapters, and edited book volumes. He was a guest editor of IEEE Signal Processing Magazine, associate editor of IEEE Transaction of Information Forensic and Security, IEEE Biometrics Compendium, and editor-in-chief of the IEEE Biometrics Council Newsletter.
\end{IEEEbiography}

\vfill

\clearpage

\setcounter{page}{1}
\setcounter{section}{0}
\setcounter{table}{0}
\setcounter{figure}{0}
\setcounter{equation}{0}

\twocolumn[ 
 \begin{@twocolumnfalse}
\begin{center}
{\Huge{Appendix: WiFaKey: Generating Cryptographic Keys from Face in the Wild}}
 \end{center}
\end{@twocolumnfalse}
]

\renewcommand{\thefigure}{A\arabic{figure}} 
\renewcommand{\thetable}{A\arabic{table}} 
\renewcommand{\thesection}{A\arabic{section}} 
\renewcommand{\thesubsection}{A\arabic{subsection}} 

\section{Face recognition models\label{sec.pretrained}}
Face recognition has been the subject of extensive study in computer vision.
It has achieved great success in the industry in recent decades due to the rapid development of GPU hardware, big data, and novel algorithms.
We briefly overview the most recent state-of-the-art face recognition work and refer readers interested in deep learning-based face recognition to survey papers such as \cite{wang2021deep} and~\cite{du2020elements}, which provide a more in-depth analysis.

The success of deep-face feature extraction techniques depends heavily on the underlying network architecture's design and the loss function's choice.
For architectures, VGGNet, GoogleNet, and ResNet~\cite{krizhevsky_imagenet_2012,simonyan_very_2014,szegedy_going_2014,he_deep_2016} are widely adopted and have a strong influence on the current state-of-the-art.
The most prevalent loss function is softmax, which attempts to promote the separability of the features.
However, the softmax loss function performs poorly for face recognition because the intra-variation could be greater than the inter-differences between identities.
Therefore, there has been much work on developing novel loss functions to make the features more separable and discriminative.
The margin-based softmax loss function has received the most attention for training face recognition models.

Arcface \cite{deng2018arcface} introduced one of the most famous margin-based softmax loss functions.
By defining the angle $\theta_{j}$ between features $\boldsymbol{f}_{i}$ and the $j$-th class center $\boldsymbol{w}_{j} \in \mathbb{R}^{d}$ as $\boldsymbol{w}_{j}^{T} \boldsymbol{f}_{i}=\left\|\boldsymbol{w}_{j}\right\|\left\|\boldsymbol{f}_{i}\right\| \cos \theta_{j}$, the loss function of Arcface can be formulated as
\begin{equation}
L_{arc}=-\frac{1}{N} \sum_{i=1}^{N} \log \frac{e^{s \cos \left(\theta_{y_{i}}+m\right)}}{e^{s \cos \left(\theta_{y_{i}}+m\right)}+\sum_{j \neq y_{i}} e^{s \cos \theta_{j}}},
\end{equation}
where $m>0$ denotes the angular margin and $s$ is the scaling parameter.
CosFace \cite{wang2018cosface} is another similar margin-based loss function, but it is additive based and defined by
\begin{equation}
L_{cos}=-\frac{1}{N} \sum_{i=1}^{N} \log \frac{e^{s \left(\cos \theta_{y_{i}}-m\right)}}{e^{s \left(\cos \theta_{y_{i}}-m\right)}+\sum_{j \neq y_{i}} e^{s \cos \theta_{j}}},
\end{equation}
where $m>0$ denotes the additive angular margin and $s$ is the scaling parameter.

However, the angular margin in Arcface and CosFace is non-adaptive and image quality-agnostic, leading to unstable intra-class distributions under image quality unconstrained scenarios.
In Magface \cite{meng2021magface}, the authors found that there is a strong correlation between the feature magnitudes and their cosine similarities with the class center; therefore, Magface is an extension of Arcface that incorporates a magnitude-aware margin and regularizer to enforce higher diversity in inter-class samples and higher similarity for intra-class samples by optimizing
\begin{equation}
\begin{aligned}
& L_{mag} = \\
& -\frac{1}{N} \sum_{i=1}^{N} [\log \frac{e^{s \cos \left(\theta_{y_{i}}+m\left(a_{i}\right)\right)}}{e^{s \cos \left(\theta_{y_{i}}+m\left(a_{i}\right)\right)}+\sum_{j \neq y_{i}} e^{s \cos \theta_{j}}} -\lambda_{g} g\left(a_{i}\right)],
\end{aligned}
\end{equation}
where $a_{i}=\left\|\boldsymbol{f}_{i}\right\|$ is the magnitude of each feature $f_{i}$, the regularizer $g\left(\cdot\right)$ is a monotonically decreasing convex function that rewards samples with large magnitudes. The hyper-parameter $\lambda_{g}$ balances the classification and regularization losses.

Adaface \cite{kim2022adaface} proposed a similar loss function with adaptive margins but based on the feature norm, as the authors observed that the feature norm could be a good indicator of image quality.
Specifically, the Adaface loss function is defined as 
\begin{equation}
L_{ada}=-\frac{1}{N} \sum_{i=1}^{N} \log \frac{e^{s \cos\left( \theta_{y_{i}}+g_{\text {angle}}\right)-g_{\text{add}}}}{e^{s \cos \left(\theta_{y_{i}}+g_{\text {angle}}\right)-g_{\text{add}}}+\sum_{j \neq y_{i}} e^{s \cos \theta_{j}}},
\end{equation}
where $g_{\text {angle}}$ and $g_{\text {add }}$ are the functions of $\widehat{\left\|\boldsymbol{f}_{i}\right\|}$ defined by 
$g_{\text {angle }}=-m \cdot \widehat{\left\|\boldsymbol{f}_{i}\right\|}, \quad g_{\text {add}}=m \cdot \widehat{\left\|\boldsymbol{f}_{i}\right\|}+m
$, where 
\begin{equation}
\widehat{\left\|\boldsymbol{f}_{i}\right\|}=\left\lfloor\frac{\left\|\boldsymbol{f}_{i}\right\|-\mu_{f}}{\sigma_{f} / h}\right\rceil,
\end{equation}
where $\mu_{f}$ and $\sigma_{f}$ are the mean and standard deviation of all $\left\|\boldsymbol{f}_{i}\right\|$ within a batch,
and $\left\lfloor\cdot\right\rceil$ refers to clipping the value between $-1$ and 1 and stopping the gradient from flowing.

On the other hand, Vision Transformer has also been adopted for face recognition.
In \cite{zhong2021facevit}, an open-set face recognition pipeline based on the standard transformer \cite{vaswani2017attention} is proposed.
The face images are split into multiple patches and input to the transformer encoder as tokens.
The cosface loss function supervises the output face image embeddings from the transformer encoder \cite{wang2018cosface}.

In summary, these works either focus on adaptive margin loss functions or emerging architectures and have demonstrated significant performance improvement under various settings, especially for unconstrained face datasets.
To use these models, pre-trained networks for Arcface, Adaface, Magface, and Face ViT are adopted in our experiments. The details of the adopted pre-trained models are summarized in the supplementary Table \ref{sup.pretrain}. 

It is worth highlighting that the employed neural-network-based feature extractors are tested on a disjoint dataset of subjects, which falls naturally into open-set recognition settings. 

\begin{table}
\centering
\caption{Summary of the pre-trained models.\label{sup.pretrain} }
\begin{tabular}{@{}cccc@{}}
\hline
\textbf{Extractor} & \textbf{Venue} & \textbf{Backbone} & \textbf{Train-set} \\
\hline
Arcface & CVPR2019 & resnet100 & ms1mv3 \\
Adaface & CVPR2022 & ir101 & webface12m \\
Magface & CVPR2021 & ir100 & ms1mv2 \\
Face ViT & arxiv & P8S8 ViT & ms1m-retinaface \\ 
\hline
\end{tabular}
\end{table}

\section{On the error correction capability \label{subsection.capbility}}
\begin{figure*}[t!]
 \centering
 \includegraphics[width=0.99\linewidth,trim=0cm 11cm 0cm 0cm, clip]{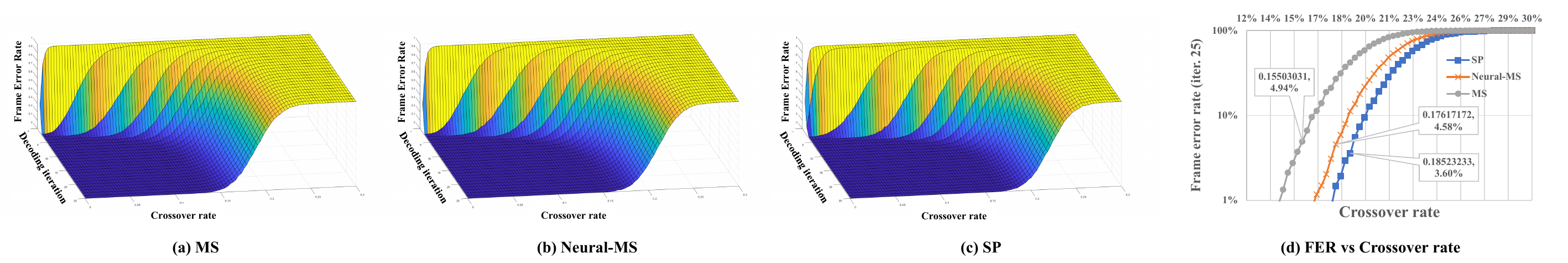} 
 \caption{Frame error rate vs.~crossover rate and number of iterations.
\label{Figure::supFERcscrossovervsiter}}
\end{figure*}

Figure \ref{Figure::supFERcscrossovervsiter} (a--c) shows the FER under different crossover rates and iterations.
Fig.~\ref{Figure::supFERcscrossovervsiter} (d) shows the FER vs.~different crossover rates after 100 iterations.
The results show that neural-MS outperforms the handcrafted MS decoder by a large margin, yet it is comparable to the SP decoder.

\section{On random masking parameter $\tau$\label{section.supmasking} }
As mentioned in the main text, $\tau$ controls the generation of masking bits and affects FMR. This section presents further results of the ablation study conducted on the FMR metric, specifically concerning the parameter $\tau$.

\begin{figure*}[t!]
 \centering
 \includegraphics[width=0.99\linewidth,trim=0cm 2cm 5cm 5.5cm, clip]{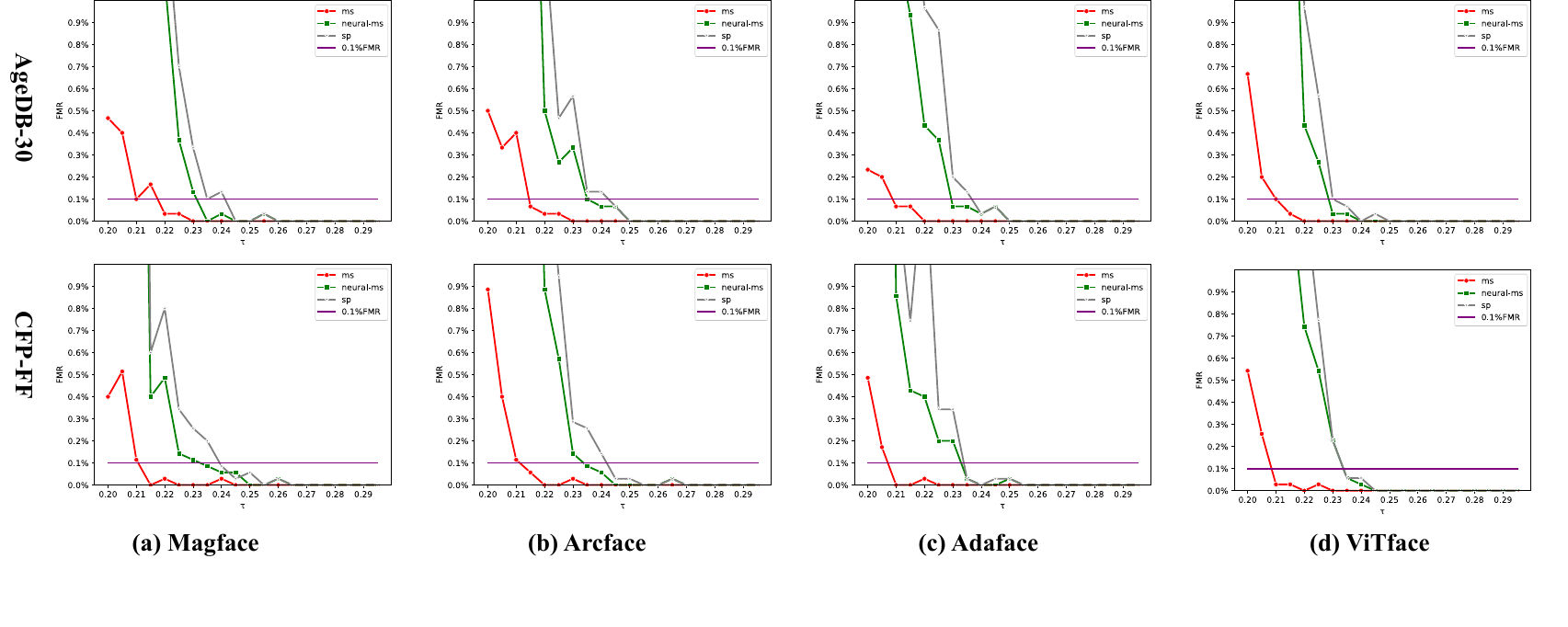} 
\vspace{-0.6cm}
\caption{FMR is mediated by $\tau$.
\label{Figure::suptauvsGMRablation}}
\vspace{-0.1cm}
\end{figure*}

Figure~\ref{Figure::suptauvsGMRablation} illustrates that MS requires a lower value of $\tau$, while SP requires a higher value to achieve a 0.1\% FMR. On the other hand, neural-MS shows a moderate performance in terms of $\tau$. Based on these observations, we fix the value of $\tau$ to 0.21 for MS, 0.235 for neural-MS, and 0.24 for SP for all subsequent experiments. Therefore, $\tau$ can be empirically determined for all datasets.

\section{Key retrieval performance of the proposed fuzzy commitment}
\begin{figure*}[t!]
 \centering
 \includegraphics[width=0.99\linewidth,trim=0cm 2cm 1cm 2cm, clip]{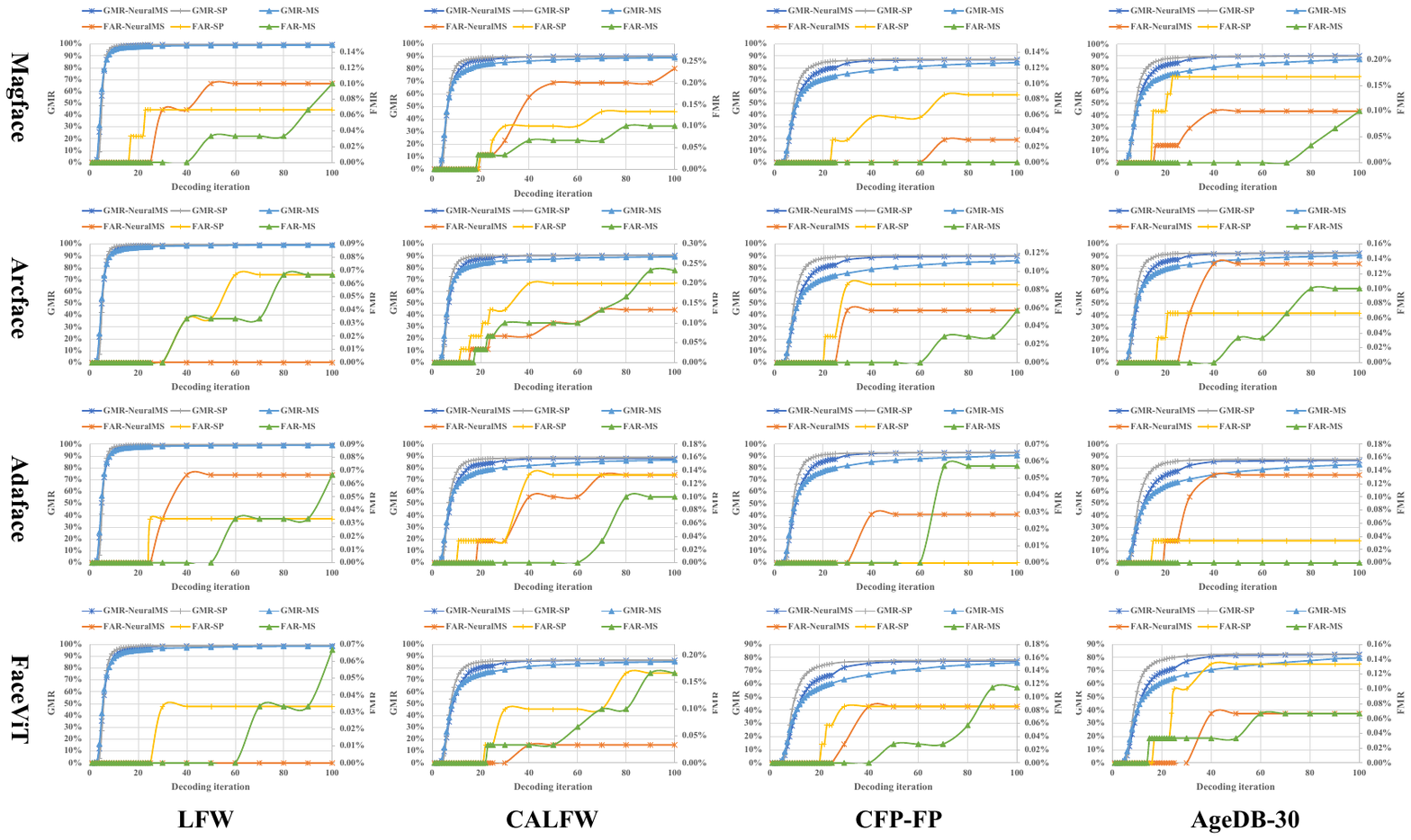} 
 \caption{Key retrieval performance in GMR and FMR vs. decoding iterations.
\label{Figure::GMRapp}}
\end{figure*}

Figure~\ref{Figure::GMRapp} presents the performance of various decoding strategies on four feature extractors. The results indicate that the handcrafted decoding method, MS, performs poorly across all datasets and feature extractors. However, the neural-MS, which uses learned weights and offset parameters, outperforms the MS decoder and performs comparably to the SP decoder.

\section{Time cost}
In this section, we evaluate the time cost of the proposed system.
The proposed fuzzy commitment mainly consists of feature extraction, feature transformation, and error correction in the decoding stage.
The error correction module is the most time-consuming, as it is an iterative decoding process.

\begin{figure}[t!]
 \centering
 \includegraphics[width=0.8\linewidth,trim=2cm 4cm 2cm 4cm, clip]{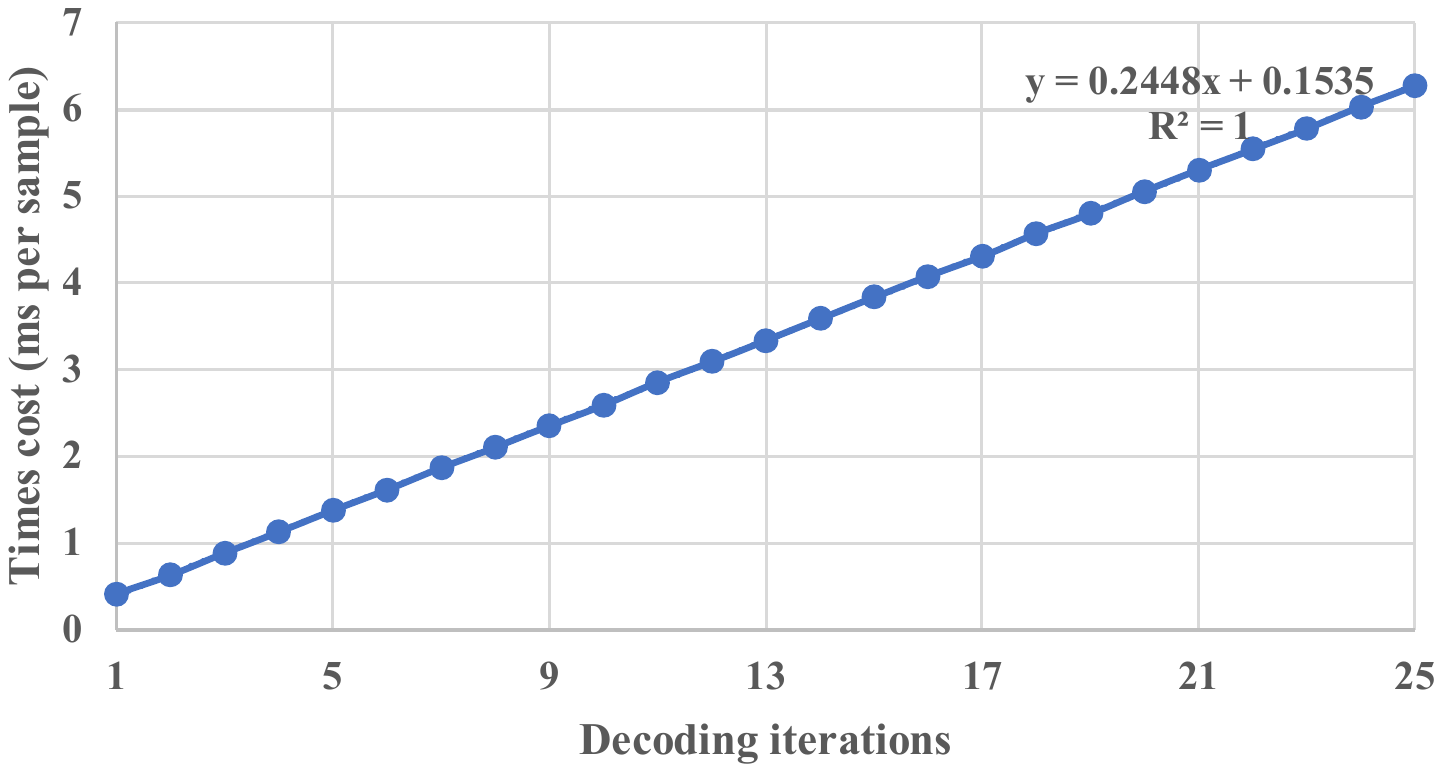} 
 \caption{Time cost.
\label{Figure::timecost}}
\end{figure}

We implemented the proposed Neural-MS decoder using TensorFlow and evaluated its time efficiency under a parallel processing pipeline with a batch size of 500 on the GTX1080Ti graphic card.
The time cost required for different numbers of iterations is shown in Fig.~\ref{Figure::timecost}.
The decoder's time cost is linear concerning the number of decoding iterations (0.2448 ms per sample and iteration).
Therefore, the time cost can be controlled carefully using the number of different iterations.

We want to stress that the LDPC is a well-researched ECC.
There are a lot of hardware-based solutions to accelerate the decoding process, such as \cite{spagnol2009hardware}; hence, the decoding efficiency can be further improved with dedicated designed hardware.

\section{Unlinkability evaluation\label{section.appendixunlink}}
We evaluate the unlinkability as specified in \cite{gomez2017general}.\cite{gomez2017general} adopted a similarity function $s$ to compute the similarity score $s=\mathrm{s}\left(t_1, t_2\right) \in \mathcal{S}$ between two templates $t_1$ and $t_2$. Given a specific similarity score $s$, the local unlinkability measure is defined as 
\begin{equation}
	\mathrm{D}_{\leftrightarrow}(s)=p\left(H_m \mid s\right)-p\left(H_{n m} \mid s\right).
\end{equation}
where
\begin{align*}
	H_{m}=&\{ \text {both templates belong to mated instances} \}\\ 
	H_{nm}=&\{ \text {both templates belong to non-mated instances} \} .
\end{align*}
The local unlinkability measure can be further relaxed as:
\begin{equation}
\mathrm{D}_{\leftrightarrow}(s)=\{
\begin{array}{ll}
	0 & \text { if } L R(s) \cdot \omega \leq 1 \\
	2 \frac{L R(s) \cdot \omega}{1+L R(s) \cdot \omega}-1 & \text { if } L R(s) \cdot \omega>1
\end{array},
\end{equation}
where
\begin{equation}
\mathcal{R}(l)=\frac{p\left(l \mid H_m\right)}{p\left(l \mid H_{n m}\right)},
\end{equation}
is the likelihood ratio between mated $\left(H_m\right)$ and non-mated $\left(H_{n m}\right)$ distributions, and $\omega=p\left(H_m\right) / p\left(H_{n m}\right)$ denotes the ratio between the prior probabilities of the mated and non-mated samples. The value of $\omega=1$, i.e, $p\left(H_m\right)=p\left(H_{n m}\right)$, is proposed as the worstcase scenario. Finally, the global measure $\mathrm{D}_{\leftrightarrow}^{s y s}$ is defined as the conditional expectation of the local measure $\mathrm{D}_{\leftrightarrow}(s)$ over all comparison scores:
\begin{equation}
\mathrm{D}_{\leftrightarrow}^{s y s}=\int p\left(s \mid H_m\right) \mathrm{D}_{\leftrightarrow}(s) \mathrm{ds} .
\end{equation}
The measure $D_{\leftrightarrow}^{s y s}$ is bounded within $[0,1]$, with $D_{\leftrightarrow}^{s y s}=1$ indicating entirely distinguishable mated and non-mated distributions, corresponding to fully linkable cae. On the other hand, $D_{\leftrightarrow}^{s y s}=0$ is achieved for fully overlapping distributions, indicating that two commitment templates derived from the same biometric trait cannot be linked.

\begin{figure*}[t!]
	\centering
	\includegraphics[width=0.99\linewidth,trim=0cm 0cm 10cm 0cm, clip]{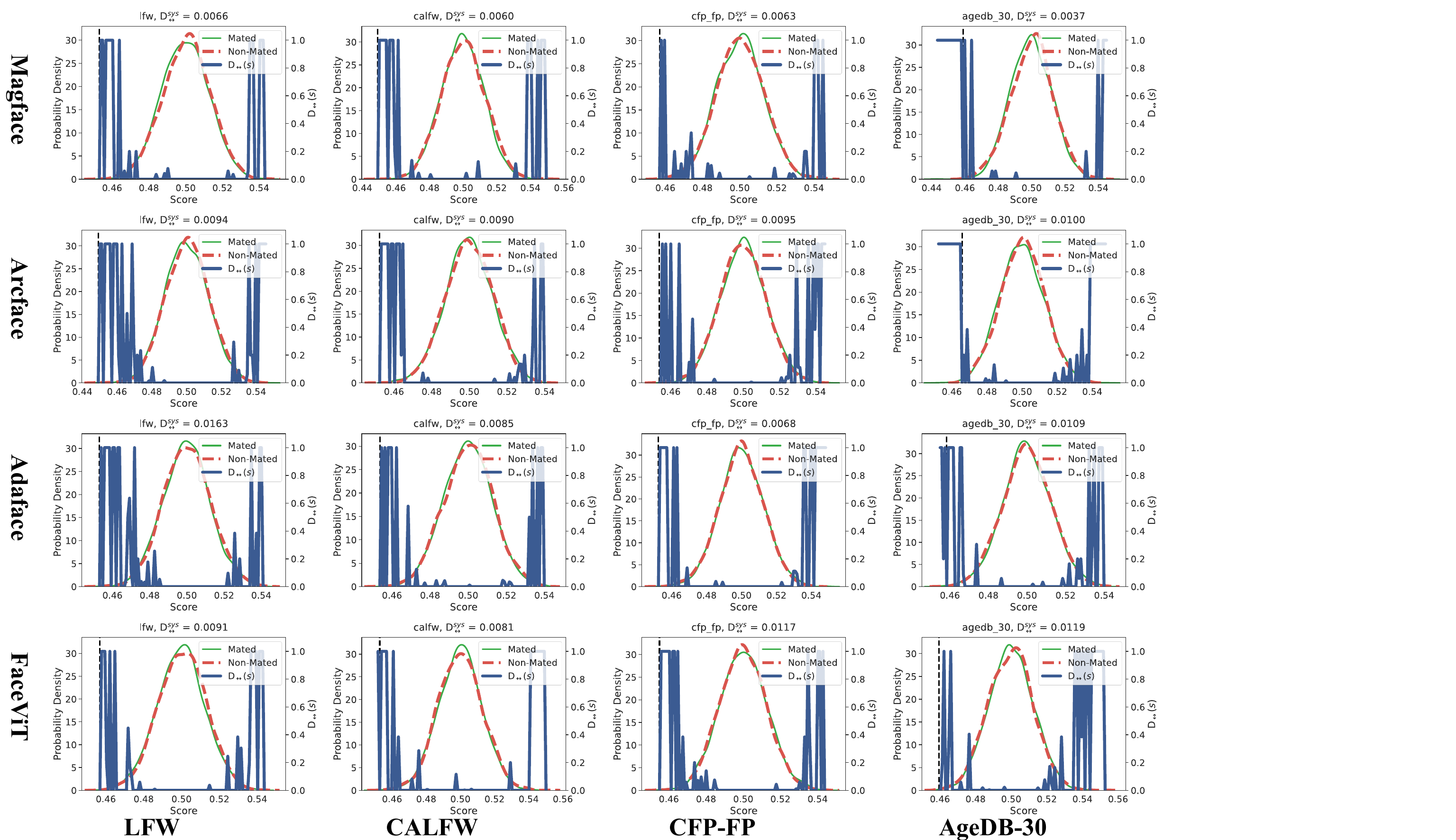} 
 \caption{Unlinkability evaluation.
\label{Figure::unlinkall}}
\end{figure*}

Using the Neural-MS parameters, the experiments were executed with Magface and on the LFW, Agedb, and CFP datasets.
A cross-matching attack can be launched by comparing the binary commitment generated from the same person in different applications (different $\mathbf{k}$).
Under this attack, the adversary is assumed to hold the commitments of the other applications.
The adversary can exploit the matching score distributions of the commitments to learn whether the commitment is from the same person. Here, we refer to the matching distance of the same subject in different applications as the mated distance.
In contrast, the matching distance from other subjects in various applications is the non-mated distance. Fig. \ref{Figure::unlinkall} gives additional experiment results on the distributions of the mated and non-mated scores.

\section{Failure case}

\begin{figure*}[t!]
	\centering
	\includegraphics[width=0.99\linewidth,trim=0cm 2cm 9cm 0cm, clip]{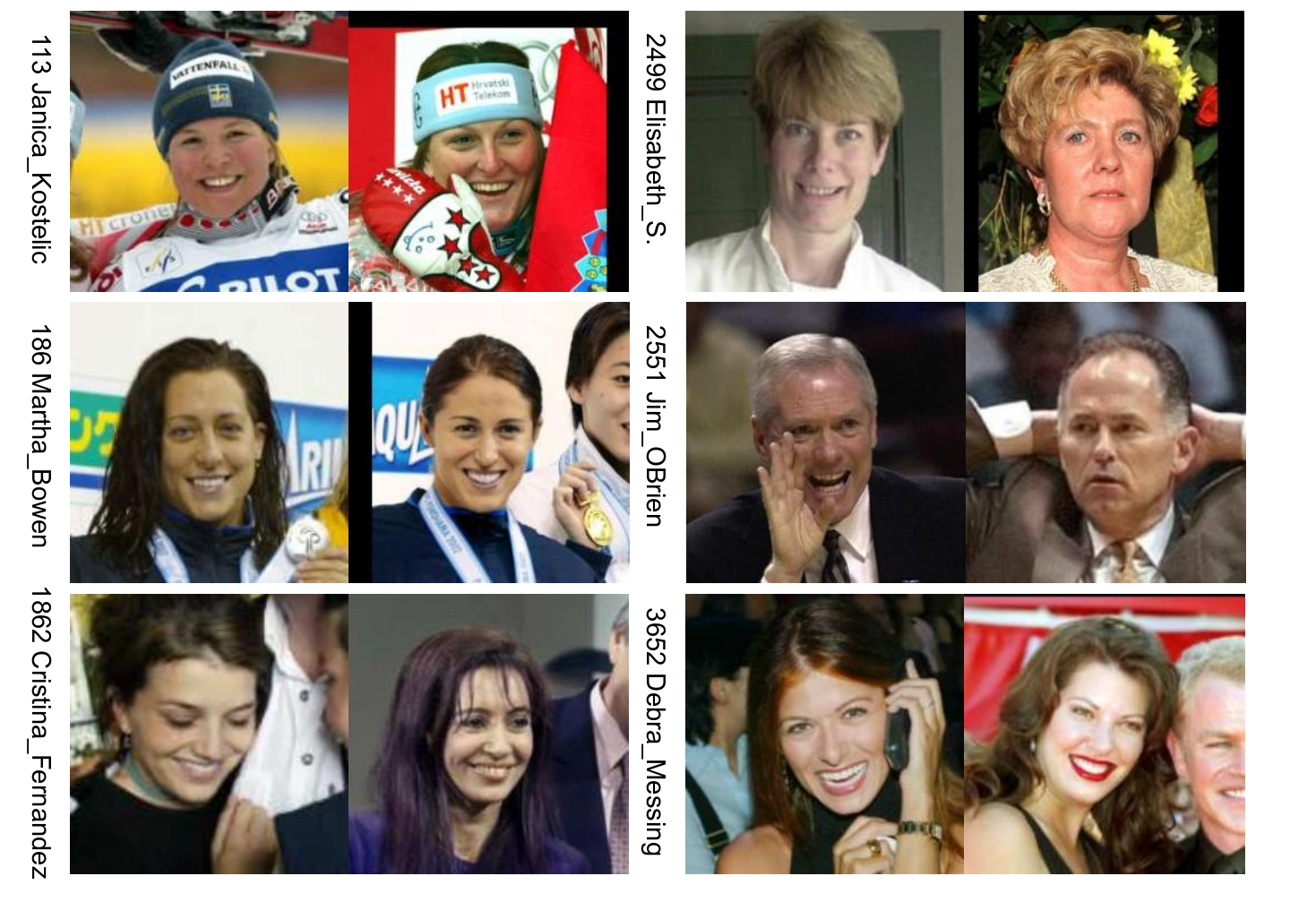} 
 \caption{Examples of failure case.
\label{Figure::failurecase}}
\end{figure*}

In our evaluation of \textit{WiFaKey}, we identified several factors that contribute to the failure in generating consistent cryptographic keys from the same individual (see Figure \ref{Figure::failurecase}):
\begin{itemize}
    \item Pose Variations: Changes in the angle or orientation of the face relative to the camera can cause discrepancies in the captured biometric features. Such pose variations lead to differences in the quantized and binarized feature representations, resulting in a higher bit error rate that exceeds the correction capability of the Neural-MS decoder.
    \item Hairstyle Changes: Alterations in hairstyle can obscure or highlight different facial features, impacting the consistency of the extracted biometric data. These changes can lead to significant variations in the feature vectors, thereby affecting the robustness of the key generation process.
    \item Lighting Conditions: Variations in lighting conditions, such as changes in color temperature, can distort the captured facial features. This inconsistency introduces noise into the measurements, leading to an increased likelihood of key generation failure.
    \item Makeup Variations: The application or removal of makeup can alter the appearance of key facial features, such as eyes, lips, and skin texture. These variations can result in substantial differences in the biometric features extracted from images taken at different times, contributing to the failure in generating consistent cryptographic keys.
\end{itemize}

These failure cases underscore the challenges associated with generating robust and reliable cryptographic keys from biometric measurements in unconstrained settings. To mitigate these issues, future work will focus on enhancing the robustness of the feature extraction and transformation pipeline. Additionally, we will explore the integration of multimodal biometric data to improve the resilience of the system. By combining facial features with other biometric modalities such as voice or iris patterns, we aim to create a more robust and secure bio-cryptosystem capable of handling a wider range of real-world scenarios.

\end{document}